\documentclass[a4paper,fleqn]{cas-dc}

\usepackage[numbers, square, comma, sort&compress]{natbib}
\usepackage{hyperref}
\hypersetup{
	colorlinks=true,       
	linkcolor=blue,        
	citecolor=blue,        
	urlcolor=blue          
}
\usepackage{doi}  
\usepackage{float}
\renewcommand{\doi}[1]{\href{https://doi.org/#1}{#1}}
\usepackage{ragged2e}
\usepackage{placeins}

\begin{document}
\let\WriteBookmarks\relax
\def\floatpagepagefraction{1}
\def\textpagefraction{.001}
\let\printorcid\relax 

\shorttitle{}

\shortauthors{M.-Q. Cheng, W.-D. Luo and H. Sun}

\title[mode=title]{\justifying 
	On-the-fly machine learning-augmented constrained AIMD 
	to design new routes from glassy carbon to quenchable amorphous diamond with low pressure and temperature%
}

\author[1]{Meng-Qi Cheng}
\author[1]{Wei-Dong Luo}
\author[1]{Hong Sun}
\cormark[1]

\ead{hsun@sjtu.edu.cn}
\affiliation[1]{organization={School of Physics and Astronomy and Key Laboratory of Artificial Structures and Quantum Control (Ministry of Education), Shanghai Jiao Tong University },
                city={Shanghai},
                postcode={200240}, 
                country={China}}

\cortext[cor1]{Corresponding author}

\begin{abstract} 
Recent advances in machine learning have enabled large-scale atomic simulations with first-principles accuracy, allowing precise modeling of disordered materials such as glassy carbon (GC). However, conventional \textit{ab initio} molecular dynamics (AIMD) cannot effectively simulate effects of anisotropic stresses, which are believed to play a key role in the transformation of GC into amorphous diamond under extreme conditions.
In this work, we present an on-the-fly machine learning-augmented constrained AIMD (ML-augmented CAIMD) approach by modifying VASP 6.3.2. Using this approach, our simulations not only reproduce experimentally observed major features of GC, but also provide restrictive conditions and microscopic analyses.  
We first demonstrate that GC has unexpectedly high plasticity, with its compression and shear strengths stiffened by large strains under ambient conditions. 
Under pressure, increasing annealing temperature promotes the formation of quenchable amorphous diamond by enhancing sp\textsuperscript{3}  preservation upon decompression, consistent with experimental observations, but this trend reverses above 2900~K due to thermal graphitization.
Under non-hydrostatic compression, we confirm that GC transforms into a superhard structure capable of sustaining large stress differences between uniaxial and confining components. This sustainable stress difference increases sharply when the confining pressure is above 40~GPa because of the continuous rising of sp\textsuperscript{3} fractions in GC induced by the applied uniaxial strains, which is absent when the confining pressure is below 30~GPa. 
Finally, we find that severe rotational shear strains applied under a pressure of 30~GPa can promote formation of a high degree sp\textsuperscript{3} hybridization (up to 80\%) at low temperatures (300$\sim$1000~K), 
which are surprisingly quenchable to ambient conditions. A hardened amorphous carbon structure retaining a substantial sp\textsuperscript{3} fraction of 64\% is obtained by decompression from a pressure of 30~GPa at 300~K after applying a rotational shear strain of 1.0, the lowest pressure and temperature ever predicted.
Our ML-augmented CAIMD algorithm not only enables designing synthesis routes for quenchable hardened amorphous carbon with unexplored conditions, but also provides a general framework to study structural transformations in disordered or defective materials under anisotropic stresses and extreme conditions.

\end{abstract}
 

\maketitle

\section{Introduction}
In recent years, machine learning potential (MLP)~\cite{Behler_PhysRevLett_98_146401_2007,Bartok_PhysRevLett_104_136403_2010,Smith_NatCommun_10_2903_2019}, which aim to accurately describe the potential energy surface of atomic configurations, have enabled large-scale molecular dynamics (MD) simulations with near \textit{ab initio} molecular dynamics (AIMD) accuracy. A well-trained machine learning force field can achieve accuracy comparable to  density functional theory (DFT)~\cite{Batzner_NatCommun_13_2453_2022,Daru_PhysRevLett_129_226001_2022}, while allowing for simulations on significantly larger spatial and temporal scales. However, conventional MLP typically require extensive training on diverse \textit{ab initio} datasets, often encompassing multiple stable phases and thermodynamic conditions for each constituent element~\cite{Pan_PhysRevB_110_224101_2024,Song_NatCommun_15_10208_2024,Wang_NatCommun_15_7607_2024}. 
This requirement poses a particular challenge for disordered materials such as glassy carbon (GC), where the structural disorder leads to highly variable bond lengths and angles that are difficult to capture using traditional training schemes based on predefined stable configurations. 
To address this limitation, the Vienna Ab Initio Simulation Package (VASP) version 6~\cite{Parrinello_PhysRevLett_45_1196_1980,Kresse_PhysRevB_47_558_1993,Kresse_PhysRevB_49_14251_1994,Kresse_PhysRevB_54_11169_1996} implements an on-the-fly machine learning force field~\cite{Jinnouchi_PhysRevB_100_014105_2019}, which adaptively updates the MLP by monitoring its deviation from DFT results during the simulation. This approach has proven particularly successful in modeling complex phase transitions such as melting. 
Furthermore, under continuous deformation such as shear or non-hydrostatic compression, the on-the-fly learning approach offers significant advantages, as the force field is adaptively refined in response to the evolving atomic environments, rather than relying on fixed, pre-trained models.
However, standard AIMD simulations in VASP (or ab initio calculations used for training traditional MLP) do not support shear strain or non-hydrostatic compression, both of which are crucial deformation patterns for GC structural transformations under extremely applied strains. 
To address this, we have modified VASP 6.3.2 to implement a machine learning-augmented constrained \textit{ab initio} molecular dynamics (ML-augmented CAIMD) framework, enabling accurate simulations of GC's structural evolution under non-hydrostatic compression and shear strain deformation.

GC, first discovered around 1960~\cite{Yamada_Nature_193_261_1962,Yamaguchi_Carbon_1_47_1963,Cowlard_JMaterSci_2_507_1967}, is an amorphous carbon material predominantly composed of sp\textsuperscript{2} bonds, which underpins its distinctive properties. Notably, GC exhibits exceptional corrosion resistance, remarkable thermal stability, excellent gas impermeability, and relatively low density (\(\sim\)1.5 g/cm\textsuperscript{3}) compared to other carbon materials. Its unique non-graphitic and highly cross-linked structure prevents graphitization until heat-treatment temperatures approach \(\sim\)3000~\textdegree C in ambient conditions ~\cite{Yamada_Nature_193_261_1962,Wang_Carbon_41_188_2003}.
Typically, GC is produced through the carbonization of thermosetting resins such as polyarylacetylene, poly (furfuryl alcohol), and phenolic resins under an inert atmosphere~\cite{Oishi_Polimeros_24_541_2014,Botelho_Carbon_39_45_2001,Oishi_ApplSurfSci_394_87_2017}. This fabrication process is well-established and has become widely industrialized, with numerous companies providing commercially available GC products~\cite{spi_supplies_2025,als_co_2025,htw_glassy_2025,merck_glassy_2025,goodfellow_catalogue_2025}. Due to its unique production method, glassy carbon exhibits notable mechanical properties, including high stiffness characterized by a Young's modulus of approximately 21~GPa~\cite{Fielda_Carbon_34_1357_1996}. Moreover, it displays a distinctive conchoidal fracture behavior similar to that of glass, a characteristic that inspired its name~\cite{Gaefke_JApplPolymSci_106_2274_2007}.
These structural and mechanical characteristics contribute to its widespread use in various scientific and industrial applications~\cite{Uskokovic_CarbonTrends_5_100116_2021,Vieira_Carbon_186_282_2022}, including ultrasensitive electrochemical sensors~\cite{Naseri_FoodChem_421_136195_2023}, catalytic systems~\cite{Cordeiro_JCatal_392_56_2020}, high-temperature oxidation-resistant environments~\cite{Murray_Carbon_167_388_2020}, antistatic packaging materials~\cite{Santos_JApplPolymSci_136_47204_2018}, wear-resistant composites~\cite{Myalski_Polymers_12_2264_2020}, and corrosion protection systems~\cite{Sure_Carbon_67_643_2014}.

Beyond its intrinsic properties and traditional applications, GC also serves as a versatile precursor for synthesizing advanced carbon materials, notably diamond-like carbon (DLC), a metastable amorphous carbon characterized by a significant percentage of sp\textsuperscript{3} bonding~\cite{Robertson_MaterSciEngR_37_129_2002}. DLC exhibits extraordinary properties akin to those of diamond, such as exceptional hardness, high elastic modulus, and excellent chemical inertness. Importantly, these advantageous characteristics occur within isotropic, disordered thin films free of grain boundaries, making DLC more economically viable compared to crystalline diamond. Consequently, DLC offers considerable benefits for various industrial and technological applications~\cite{Erdemir_JPhysDApplPhys_39_R311_2006,Hauert_SurfCoatTechnol_233_119_2013,Duan_JMaterChemC_11_5585_2023,Bai_JNonCrystSolids_443_8_2016}. The sp\textsuperscript{3} bonding content in carbon-based DLC critically determines its mechanical hardness, chemical and electrochemical inertness, and wide electronic band gap~\cite{Xu_PhilosMagB_76_351_1997}. Owing to this critical role, the classification of carbon-based DLC is commonly based on its sp\textsuperscript{3} bonding fraction. Pure carbon-based DLC is typically classified  by ranging its sp\textsuperscript{3} bonding percentage into: (i) Amorphous carbon (a-C), possessing relatively low sp\textsuperscript{3} content less than \(<\) 80\%. (ii) Tetrahedral amorphous carbon (ta-C), characterized by a fairly high sp\textsuperscript{3} content ranging approximately between 80\textendash 88\%~\cite{McKenzie_RepProgPhys_59_1611_1996,Fallon_PhysRevB_48_4777_1993,Pharr_ApplPhysLett_68_779_1996,Tan_JPhysChemC_124_5489_2020}, which exhibits properties more closely resemble those of crystalline diamond because of its higher sp\textsuperscript{3} hybridization density~\cite{Robertson_MaterSciEngR_37_129_2002,Klein_Carbon_107_536_2016,Seok_SciRep_8_13521_2018,Zhao_JApplPhys_135_065304_2024}. (iii) Amorphous diamond (a-D), a recently synthesized novel carbon allotropes including nearly full sp\textsuperscript{3}-bonded amorphous structures utilizing GC as a precursor under extreme conditions, such as high-pressure and high-temperature or high-pressure only environments, where nanocrystalline cubic diamond or hexagonal diamond (lonsdaleite) sometimes appear as byproducts~\cite{Tan_JPhysChemC_124_5489_2020,Lin_PhysRevLett_107_175504_2011,Shiell_SciRep_6_37232_2016,Zeng_NatCommun_8_322_2017,McCulloch_Small_16_2004695_2020,Huang_Carbon_219_118763_2024}.

a-D is an isotropic, disordered structure that maintains hardness comparable to diamond, offering superior isotropic mechanical properties. Lin et al. (2011) first identified this superhard carbon allotrope by compressing GC above 40~GPa at room temperature, measuring a remarkable strength of approximately 130~GPa at a confining pressure of 60~GPa~\cite{Lin_PhysRevLett_107_175504_2011}. However, experimental findings at room temperature have revealed notable discrepancies. For instance, Solopova et al. (2013) reported no structural transformation even at pressures as high as 60~GPa~\cite{Solopova_ApplPhysLett_102_121909_2013}. In contrast, Yao et al. (2014) detected a structural transformation at approximately 33~GPa~\cite{Yao_ApplPhysLett_104_021916_2014}, aligning with earlier observations~\cite{Lin_PhysRevLett_107_175504_2011}, and recorded a yield strength of 120~GPa under a confining pressure of 62~GPa, comparable to that of diamond under ambient conditions. Yao et al. attributed this discrepancy to differences in experimental pressure conditions, proposing that shear stress might significantly influence the structural  transformation~\cite{Yao_ApplPhysLett_104_021916_2014,Yao_ApplPhysLett_111_101901_2017}. However, the superhard phase of GC under high pressures observed in the above experiments~\cite{Lin_PhysRevLett_107_175504_2011,Yao_ApplPhysLett_104_021916_2014,Yao_ApplPhysLett_111_101901_2017} are not quenchable to normal ambient conditions where the rich sp\textsuperscript{3} bonds disappear upon compression releasing, though permanent graphitization and densification of GC structures are identified if unixial compression pressures are higher than 45 GPa at room temperature~\cite{Shiell_PhysRevLett_120_215701_2018}. Quenchable a-D (sp\textsuperscript{3} fraction \(\sim\) 100\%) or a-C (sp\textsuperscript{3} fraction \(\sim\) 50\%) can be synthesized from GC with high pressure and high temperature (HPHT) conditions (\(p\sim\)~50 GPa, \(T\sim\)~1800 K)~\cite{Zeng_NatCommun_8_322_2017}, or high pressure and thermal annealing (\(p\sim\)~58 GPa, \(T\sim\)~728 K) during which most of the sp\textsuperscript{3} bonds are adjusted to low energy stable configurations~\cite{Zeng_PhysRevB_109_214113_2024}. One of the aims in the present work is to find new pathways to synthesize quenchable sp\textsuperscript{3}-rich bonding amorphous carbon structures from GC with much reduced pressure and temperature. Recently, severe plastic deformations (SPDs) produced by continuous shearing in rotational diamond anvil cell (RDAC) have been applied to various materials as a new controlling parameter to study their phase transformation and structural evolution, with the latest reviews in Ref.\cite{Levitas_MaterialsTrans_64_1866_2023,Edalati_MaterialsTrans_65_466_2024}. Particularly, graphite, graphene nanoplatelet, carbon nanotube and C\textsubscript{60} fullerene, where sp\textsuperscript{2} bonding dominates the structures, are experimentally investigated subject to severe plastic shear deformation in RDAC~\cite{Gao_Carbon_146_364_2019,Yang_Carbon_232_119802_2025} , where stable sp\textsuperscript{3} carbon bonds appear in nano-diamond form under pressures even lower than 1 GPa at room temperature. By combining effects of temperature, pressure and SPDs, one can design new routes to synthesize the superhard phase of GC with easier experimental conditions.

Although experimental findings underscore the pivotal role of shear in driving the transformation of GC into advanced carbon materials, comprehensive theoretical investigations into shear or other non-hydrostatic compression mechanisms remain scarce. Most of the calculations reported are restricted to simple uniaxial compression where the lateral cross sections of GC samples remain unchanged or 0~K temperature is assumed~\cite{Shiell_PhysRevLett_120_215701_2018,Huang_Carbon_219_118763_2024,Zeng_PhysRevMater_3_033608_2019}. But, to the best of our knowledge, effects of severe large shear deformations from RDAC on the structural evolution of GC have not been investigated so far. Such studies 
can systematically explore various conditions of temperature, pressure and SPDs, thereby identify precise and industrially accessible critical conditions for transforming GC into a-D or other carbon allotropes. The complexity of addressing this theoretical challenge primarily stems from the following two facts: (i) The inherently disordered nature of GC~\cite{Jenkins_Nature_231_175_1971,Yoshida_Carbon_29_1107_1991,Harris_PhilosMagA_84_3159_2004,Jurkiewicz_JApplCrystallogr_50_36_2017} which complicates the construction of accurate atomic models capable of faithfully replicating experimental observations; (ii) Lack of DFT programs, independent of artificially adjustable parameters, to simulate structural evolution of large GC supercells under combined influences of temperature, pressure and SPDs.

Compared to well-established experimental fabrication techniques~\cite{Oishi_Polimeros_24_541_2014,Botelho_Carbon_39_45_2001,Oishi_ApplSurfSci_394_87_2017,spi_supplies_2025,als_co_2025,htw_glassy_2025,merck_glassy_2025,goodfellow_catalogue_2025}, only a few theoretical approaches have been developed to construct GC structures. One common approach simulates thermal treatment of precursors containing C, H, and O at high temperatures, followed by controlled cooling~\cite{MontgomeryWalsh_Carbon_184_627_2021}. However, periodic boundary conditions prevent species like H and O from completely escaping, leading to impurity retention in the resulting models. Consequently, this approach mainly investigates precursor-to-GC transformations rather than generating pure GC structures. Furthermore, due to the substantial atomic systems required to match experimental scales, Classical Molecular Dynamics (CMD) simulations with predefined potentials are typically employed.

Another theoretical approach constructs GC models by assembling multiple carbon material fragments within a cubic box~\cite{Wen_PhysRevB_98_014103_2018,Jiang_SciRep_3_1877_2013}. This method produces relatively small GC structures suitable for first-principles calculations at 0~K. However, as the structural units originate from existing carbon materials, the generated GC structures may differ significantly from experimentally synthesized GC samples.

The third method, known as liquid quenching, involves melting carbon atoms at high temperatures, followed by rapidly quenching and subsequently annealing at specified temperatures~\cite{Tomas_Carbon_109_681_2016,Hossain_Carbon_183_940_2021,Yeh_Carbon_234_120006_2025}. Tomas et al. utilized CMD simulations with various carbon potentials, observing inconsistencies among results obtained using different potentials~\cite{Tomas_Carbon_109_681_2016}, highlighting CMD simulation limitations. Thus, this method is frequently applied alongside experimental validation to ensure accuracy~\cite{Shiell_PhysRevLett_120_215701_2018,Huang_Carbon_219_118763_2024}. We will show below that the calculated results of GC by CMD under very high pressure or applied large strains deviate from those obtained by our on-the-fly ML-augmented CAIMD. Zeng et al.~\cite{Zeng_PhysRevMater_3_033608_2019,Tan_JPhysChemC_124_5489_2020} employed this liquid quenching approach with AIMD simulations using VASP~\cite{Parrinello_PhysRevLett_45_1196_1980,Kresse_PhysRevB_47_558_1993,Kresse_PhysRevB_49_14251_1994,Kresse_PhysRevB_54_11169_1996}, successfully reproducing experimental observations with an GC supercell containing 1024 atoms. However, due to the immense computational cost, they employed AIMD only for structure generation, with subsequent calculations for uniaxial compression performed at 0~K using first-principles calculations.

In this study, we develop an on-the-fly ML-augmented CAIMD simulation program to investigate the structural evolution of
GC as a function of pressure, temperature, uniaxial compression and severe rotational shear strains. Our simulations not only successfully reproduced and complemented several experimentally observed structural transformations of GC under various temperature and pressure conditions~\cite{Shiell_PhysRevLett_120_215701_2018,Hu_JMateriomics_7_177_2021,Zeng_NatCommun_8_322_2017}, but also provide restrictive conditions and microscopic analyses.  
We first demonstrate the 
 highly plastic and strain-stiffening behaviors of GC under large tensile, compression, and shear deformation in ambient conditions, and reveal their underlying atomistic mechanisms using ML-augmented CAIMD simulations.
 Under high pressure, we find that annealing enhances sp\textsuperscript{3} preservation upon decompression, promoting the formation of quenchable a-D. This effect enhances with increasing temperature but critically reverses above 2900~K due to thermal graphitization. However, we show that by applying pressure at room temperature before heating, graphitization can be averted, allowing annealing at 3000~K to successfully yield an quenchable a-D structure. This strategy effectively extends the accessible temperature window for sp\textsuperscript{3}-rich carbon synthesis by avoiding high temperature graphitization. 
In simulations involving pressure application at room temperature followed by heating, we also identified a critical threshold (30--40~GPa) above which sp\textsuperscript{3} bonding increases markedly with temperature, in good agreement with experimental values of 33 or 40~GPa~\cite{Lin_PhysRevLett_107_175504_2011,Yao_ApplPhysLett_104_021916_2014}.
We also find that, under non-hydrostatic compression, GC undergoes a transition into a superhard phase capable of maintaining substantial stress differences between uniaxial and confining directions. This stress difference increases significantly when the confining pressure is above 40~GPa, driven by a steady increase in sp\textsuperscript{3} content induced by uniaxial strain, whereas no such effect is observed below 30~GPa.
Finally, we examined the role of rotational shear strains at fixed pressures. Our results show that shear strain can induce substantial sp\textsuperscript{3} bonds, reaching up to 80\% at relatively low temperatures, even at pressures slightly below the critical threshold (e.g., 30~GPa).
Moreover, compared to hydrostatic pressure alone, the sp\textsuperscript{3} bonding generated via shear strain exhibits markedly enhanced structural stability. This stability is attributed to the distribution of more C--C bonds with lengths closer to the ideal sp\textsuperscript{3} bond length of cubic diamond under the same conditions, which facilitates the preservation of high sp\textsuperscript{3} content even after pressure and strain are released at room temperature. As a representative case, applying a rotational shear strain of 1.0 under 30~GPa at 300~K yields an amorphous hardened carbon structure with 64\% sp\textsuperscript{3} content, the lowest pressure and temperature conditions predicted so far. This demonstrate the potential of our simulate scheme to model structural evolution under extreme anisotropic loading and to guide the design of synthesis routes for amorphous hardened carbon structures.

\section{Methods of Calculations}
Our atomistic simulations were performed using the on-the-fly ML-augmented AIMD framework implemented in VASP~\cite{Parrinello_PhysRevLett_45_1196_1980,Kresse_PhysRevB_47_558_1993,Kresse_PhysRevB_49_14251_1994,Kresse_PhysRevB_54_11169_1996,Jinnouchi_PhysRevB_100_014105_2019}, where the ML force field is continuously checked and updated with AIMD data, enabling more accurate modeling of phase transitions and sustained deformation~\cite{Jinnouchi_PhysRevB_100_014105_2019}. We employed the projector augmented-wave method~\cite{Kresse_PhysRevB_59_1758_1999} and the Perdew-Burke-Ernzerhof exchange-correlation functional optimized for solids and surfaces (PBEsol)~\cite{Perdew_PhysRevLett_100_136406_2008,Jiang_SciRep_3_1877_2013} within the generalized gradient approximation~\cite{Perdew_PhysRevLett_77_3865_1996},  with the DFT-D3 correction to treat the Van der Waals interaction. All simulations utilized a plane-wave energy cutoff of 500~eV, a timestep of 1~fs, and a \(1 \times 1 \times 1\) Monkhorst-Pack k-point grid~\cite{Monkhorst_PhysRevB_13_5188_1976}.

\begin{figure}
	\centering
	\includegraphics[width=0.45\textwidth]{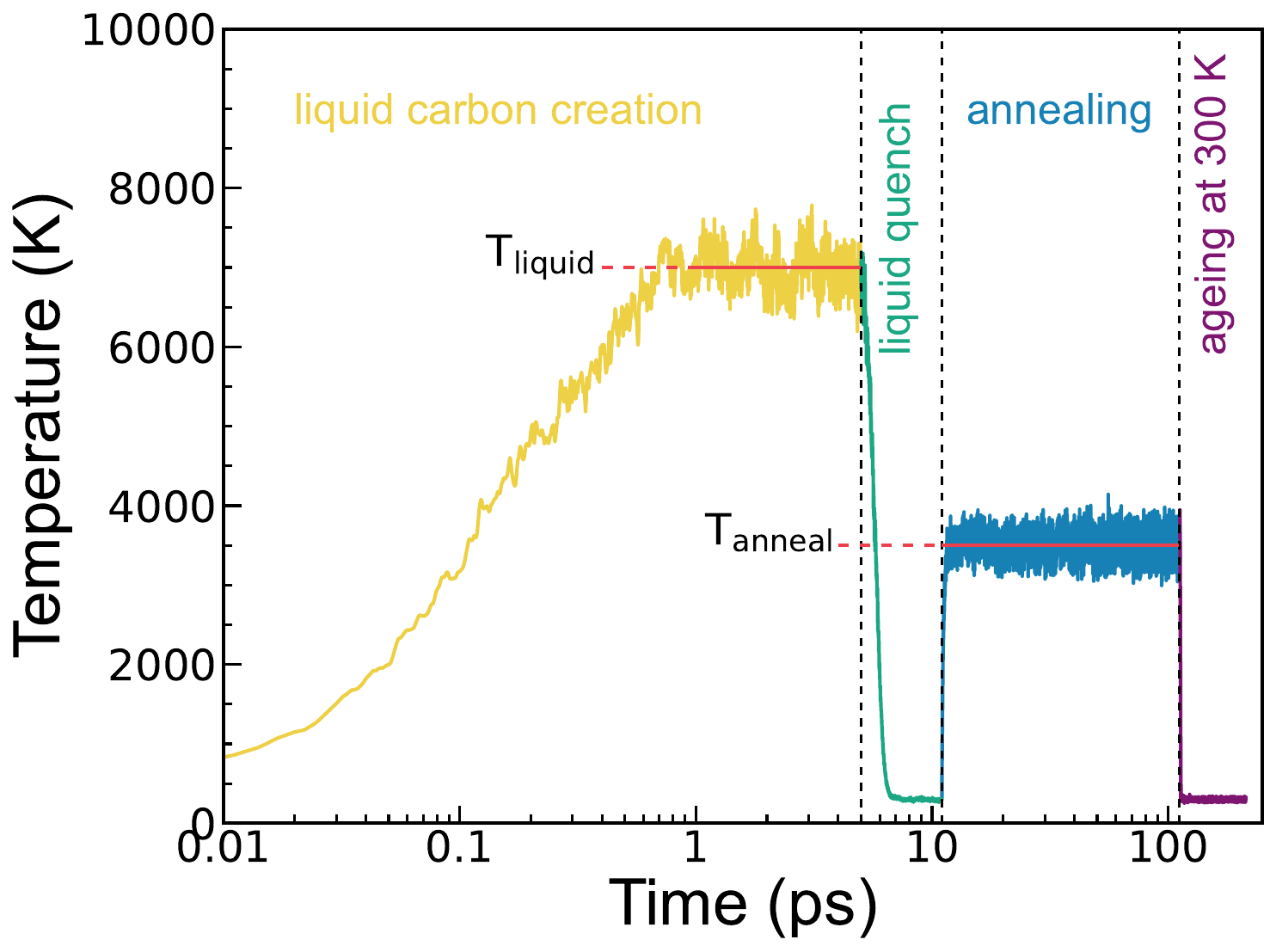}
	\caption{The temperature evolution as a function of AIMD time throughout the entire simulation process of GC preparation. Initially, the structure was melted at T\textsubscript{liquid} (7000~K), followed by a rapid quenching to 300~K and subsequent annealing at T\textsubscript{anneal} (3500~K). During these stages, an \textit{NVT} ensemble was employed to maintain a constant volume (and thus a fixed density). Finally, the system was cooled to 300~K within an \textit{NpT} ensemble, allowing for structural relaxation which induces a small density variation due to volume changes.}
	\label{FIG:1}
\end{figure}

The initial atomic configuration of GC was prepared through the liquid quenching method (Fig.~\ref{FIG:1}), which is widely adopted in previous computational studies of GC~\cite{Tan_JPhysChemC_124_5489_2020,Tomas_Carbon_109_681_2016,Hossain_Carbon_183_940_2021,Yeh_Carbon_234_120006_2025,Shiell_PhysRevLett_120_215701_2018,Zeng_PhysRevMater_3_033608_2019,Huang_Carbon_219_118763_2024}. The initial structure, comprising 500 carbon atoms with a density of 1.5~g/cm\textsuperscript{3}, was generated by randomly clipping multiple fullerene fragments~\cite{Wen_PhysRevB_98_014103_2018}, and its dynamical stability at room temperature (300~K) was verified. 
Following the liquid quenching protocol, we conducted on-the-fly AIMD simulations within the \textit{NVT} ensemble. Initially, the structure was melted entirely at 7000~K (T\textsubscript{liquid}), characterized by a linear increase in the mean-squared displacement (MSD) over time~\cite{Tomas_Carbon_109_681_2016}. After equilibration at this high temperature for 5~ps, the system was rapidly quenched~\cite{Marks_PhysRevB_56_2441_1997} to 300~K over 1~ps and subsequently equilibrated for an additional 4~ps to form an amorphous carbon solid. 
The resulting amorphous phase underwent annealing at 3500~K (T\textsubscript{anneal}) for 100~ps, a temperature high enough to facilitate structural evolution yet below the melting threshold, as determined from MSD analysis~\cite{Tomas_Carbon_109_681_2016}. Finally, the system was cooled to 300~K and aged for 100~ps within the \textit{NpT} ensemble~\cite{Parrinello_PhysRevLett_45_1196_1980,Parrinello_JApplPhys_52_7182_1981}, after which subsequent simulations were conducted. Short 1~ps heating and cooling periods were consistently applied prior to annealing and aging stages.

\begin{figure}
	\centering
	\includegraphics[width=0.45\textwidth]{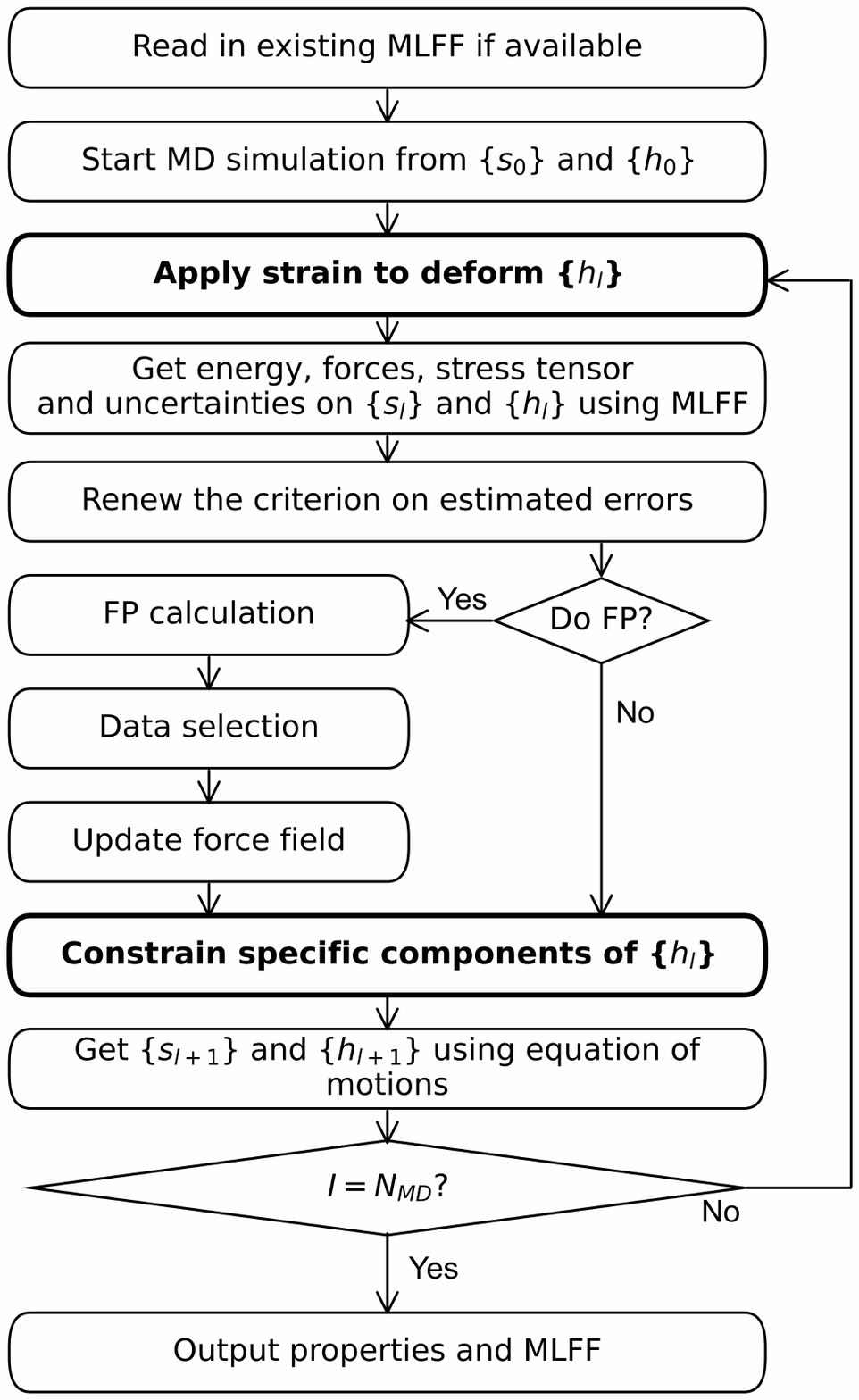}
	\caption{Flowchart illustrating the modified VASP workflow for constrained deformation simulations under the \textit{NpT} ensemble coupled with on-the-fly MLFF generation. The steps highlighted in Boldface (\textbf{Apply strain} and \textbf{Constrain}) represent our newly introduced procedures to impose and sustain applied anisotropic strains (tensile, compression, or shear) during simulations. \{s$_l$\} describes atomic positions of all atoms in the supercell at the $l$-th CAIMD step. \{h$_l$\} is a matrix with nine components from the three lattice vectors of the supercell at the $l$-th CAIMD step given in SM. FP is the abbreviation for first-principles (calculation). N$_{\rm MD}$ is the total CAIMD steps to be performed.}
	\label{FIG:2}
\end{figure}

For the \textit{NpT} AIMD simulations, VASP employs the Parrinello-Rahman algorithm~\cite{Parrinello_PhysRevLett_45_1196_1980,Parrinello_JApplPhys_52_7182_1981}, but originally, it considers only hydrostatic pressure. To facilitate constrained deformation under the \textit{NpT} ensemble, we modified the VASP AIMD source code into CAIMD by limiting some of the equations of motions for the unit vectors in the Parrinello and Rahman method~\cite{Parrinello_JApplPhys_52_7182_1981}. The theoretical details can be found in the Supplementary Material (SM) (also see the discussion below). This CAIMD has been used by our group to investigate the effects of shear and tensile strains on the melting temperatures of diamond and TaC~\cite{Wen_Carbon_155_361_2019,Yang_PhysRevB_107_104101_2023}. The exact CAIMD can simulate structural deformations with supercells containing about one hundred atoms due to the hugeous amount of first-principles calculations. However, it can integrate with the VASP's on-the-fly machine learning force field (MLFF) naturally~\cite{Jinnouchi_PhysRevB_100_014105_2019}, where the MLFF is continuously checked and updated with CAIMD data during structural deformations. This coupling of CAIMD with on-the-fly MLFF
enables efficient simulations of constrained material deformations while preserving the accuracy of first-principles calculations. An estimated computing run-time is given at the end of this section.
Our modifications includes primarily two additional steps in the original MLFF generation workflow~\cite{Jinnouchi_PhysRevB_100_014105_2019}, as illustrated in Fig.~\ref{FIG:2}. Specifically, before evaluating the energy, force, and stress tensors, either via MLFF or first-principles methods, we insert an explicit \textbf{Apply strain} step. This step incrementally adjusts the lattice vectors of GC supercells, according to the type of anisotropic stress applied (tensile, compression, or shear). Throughout all simulations, various types of strains are applied through a constant strain rate of 0.001~ps\(^{-1}\)~\cite{Wen_Carbon_155_361_2019,Yang_PhysRevB_107_104101_2023,Cheng_PhysRevMater_8_113604_2024}, which linearly increases the applied strains as the CAIMD calculation steps develop.
The imposed strain drives the material away from equilibrium, resulting in deviations of the stress tensor components from their equilibrium values (near-zero if without hydrostatic pressure). Therefore, prior to updating atomic positions and lattice parameters in the next integration step according to the MD equations of motion, we constrain the changes of the specific components of lattice vectors which are adjusted by the given applied strain. One example of such constrain is given in Eq.(S6) of SM for the uniaxial compression along the $z$ axis.  Without this constrain, the strained material would inadvertently relax back to equilibrium during the dynamical update. We term this process as a \textbf{Constrain} step. 

\begin{figure}
	\centering
	\includegraphics[width=0.45\textwidth]{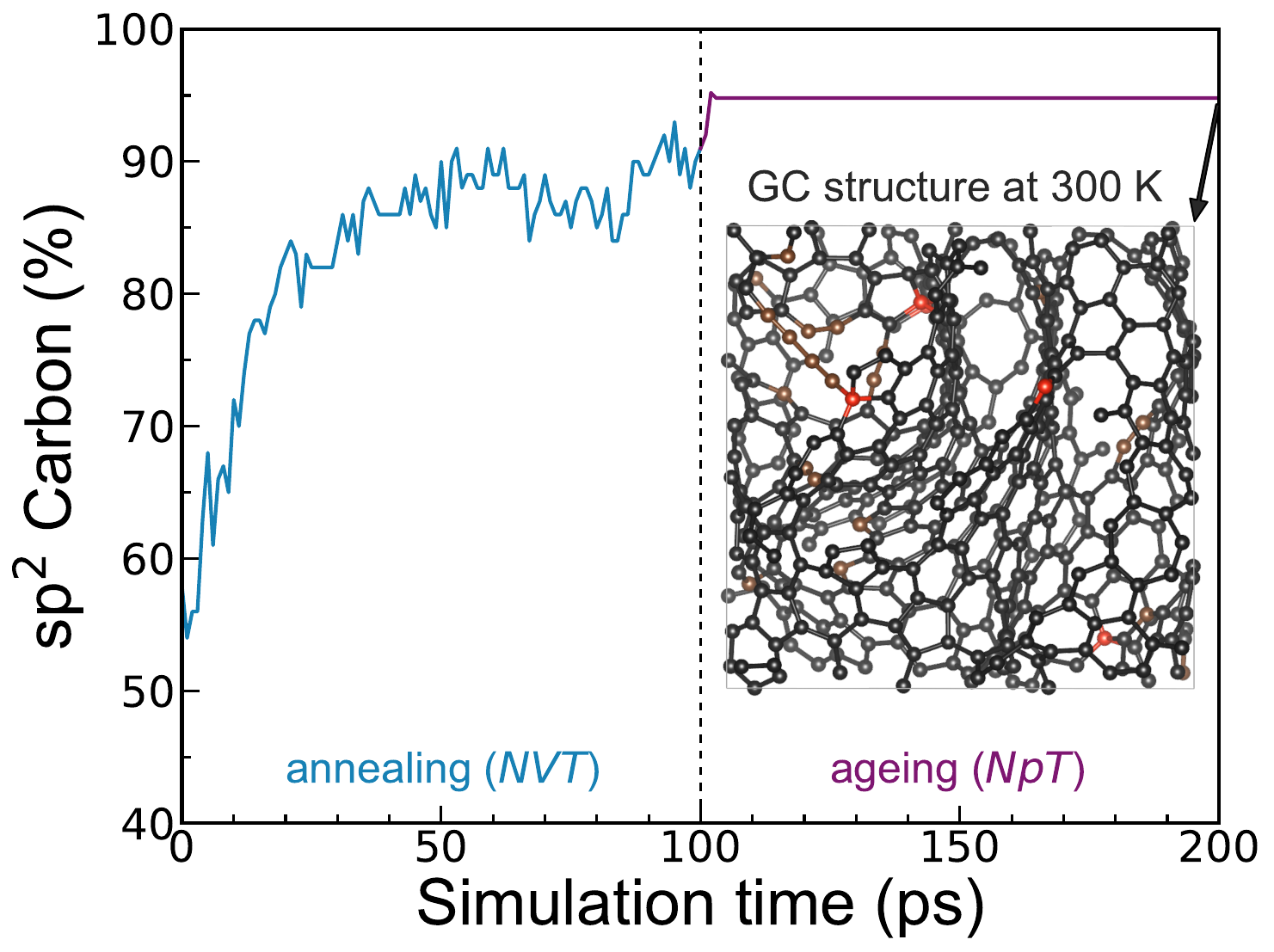}
	\caption{Evolution of the percentage of sp\textsuperscript{2} bonds during the annealing (100~ps) and aging (100~ps) processes as a function of simulation time. Data was sampled at 1~ps intervals. The inset shows the final GC structure at 201~ps (including an extra 1~ps cooling stage before aging). 
	Brown-colored atoms represent sp-bonded carbon, gray-colored atoms indicate sp\textsuperscript{2}-bonded carbon, and red-colored atoms correspond to sp\textsuperscript{3}-bonded carbon.}
	\label{FIG:3}
\end{figure}

Among various types of anisotropic stresses, we primarily focus on uniaxial compression and shear deformation, which has been repeatedly emphasized as crucial in previous studies~\cite{McCulloch_Small_16_2004695_2020,Huang_Carbon_219_118763_2024,Yao_ApplPhysLett_104_021916_2014}. The uniaxial compression, and similarly linear shear, deformation process carried out in CAIMD has been presented in SM.
Here we mainly discuss how to simulate rotational shear deformation using CAIMD at finite temperatures similar to that imposed by RDAC~\cite{Ma_JPhysChemSolids_67_2083_2006,Ciezak_RevSciInstrum_82_073905_2011} commonly employed in experimental studies.  
The rotational shear is carried out on a plane normal to the $z$ axis. We first apply a linear shear strain of 0.2 in any direction (according to the strain rate of 0.001~ps\(^{-1}\) and timestep of 1~fs, it needs 200,000 CAIMD steps) to the GC supercell to move it away from the equilibrium structure, followed by performing 16 consecutive linear shear operations, each with the shear direction rotated by 22.5\textdegree, which ultimately returns to the initial orientation before rotation, as sketched in Fig.~S1 of SM. In each of these 16 shear directions, a linear shear strain of 0.2 is applied with the same strain rate, which requires a total of 3,400,000 CAIMD steps (3400 ps) to complete the full rotation cycle. 
Naturally, all the on-the-fly ML-augmented CAIMD simulations were performed under the same criterion conditions similar to those described in previous studies~\cite{Wen_Carbon_155_361_2019,Yang_PhysRevB_107_104101_2023,Cheng_PhysRevMater_8_113604_2024}. Specifically for the shear deformation, the shear strain \(\epsilon_{zx}=\dot{\varepsilon} \times t\) increases as CAIMD steps go on, which results in a corresponding non-zero shear stress $\sigma_{zx}$ due to the imposed constrain in CAIMD, while all the other components of stresses and forces on atoms satisfy the criterion: (i) the mean values (averaged over 2000 CAIMD steps) of the axial stress components \(\sigma_{xx}\), \(\sigma_{yy}\), and \(\sigma_{zz}\) equal to the applied hydrostatic pressure; (ii) the mean values of the shear stress components \(\sigma_{xy}\) and \(\sigma_{yz}\) approach zero (<0.1~GPa); and (iii) the average force on each atom (averaged over 2000 CAIMD steps) is negligible (<0.1~eV/Å). A detailed comparison between the on-the-fly ML-augmented CAIMD and exact CAIMD is provided in Fig.~S2 of SM. We found that the on-the-fly ML-augmented CAIMD improves computational efficiency by 95$\sim$99\% while maintaining an error of only 2.3\% compared to the exact CAIMD calculations.  A full circle rotation shear described above involving 3,400,000 CAIMD steps (3400~ps) for a GC supercell of 500 atoms (see Fig.~\ref{FIG:3}) can be completed within 3$\sim$6 days with 128 CPU cores depending on the temperature and pressure considered in our study.

\section{Results and discussions}
Firstly, we investigate whether the produced structure conforms to the characteristic features of GC, specifically its low density (1.5 g/cm\textsuperscript{3}) and high percentage of sp\textsuperscript{2} bonds.
In Fig.~\ref{FIG:3}, we present the evolution of the percentage of sp\textsuperscript{2} bonds (every 1~ps) during the annealing (100~ps) and aging (100~ps) processes. It is evident that extended annealing leads to an increase in the percentage of sp\textsuperscript{2}-hybridization, with approximately 50~ps sufficient to achieve the maximum sp\textsuperscript{2}-hybridization content in our structure. Upon cooling to 300~K (1~ps) within an \textit{NpT} ensemble before aging, the sp\textsuperscript{2}-hybridization percentage further increases, eventually stabilizing at a high constant value (\(\sim\)95\%). Although slight volume fluctuations occur during the aging period, influencing the density, the average density during the final 50~ps remains relatively stable at approximately 1.50~g/cm\textsuperscript{3}, reaching a final average of 1.54~g/cm\textsuperscript{3}. To explore the effects of higher temperatures, we systematically increase the temperature from 300~K to 1000~K, 2000~K, 2900~K, and 3000~K, with the resulting average densities and sp\textsuperscript{2}-hybridization percentages summarized in Table~\ref{tab1}. 
Each temperature condition is simulated for 100~ps using the on-the-fly ML-augmented AIMD (hereafter referred to simply as simulation), and average values are obtained from the last 50~ps following equilibration.

\begin{figure*}
	\centering
	\includegraphics[width=0.9\textwidth]{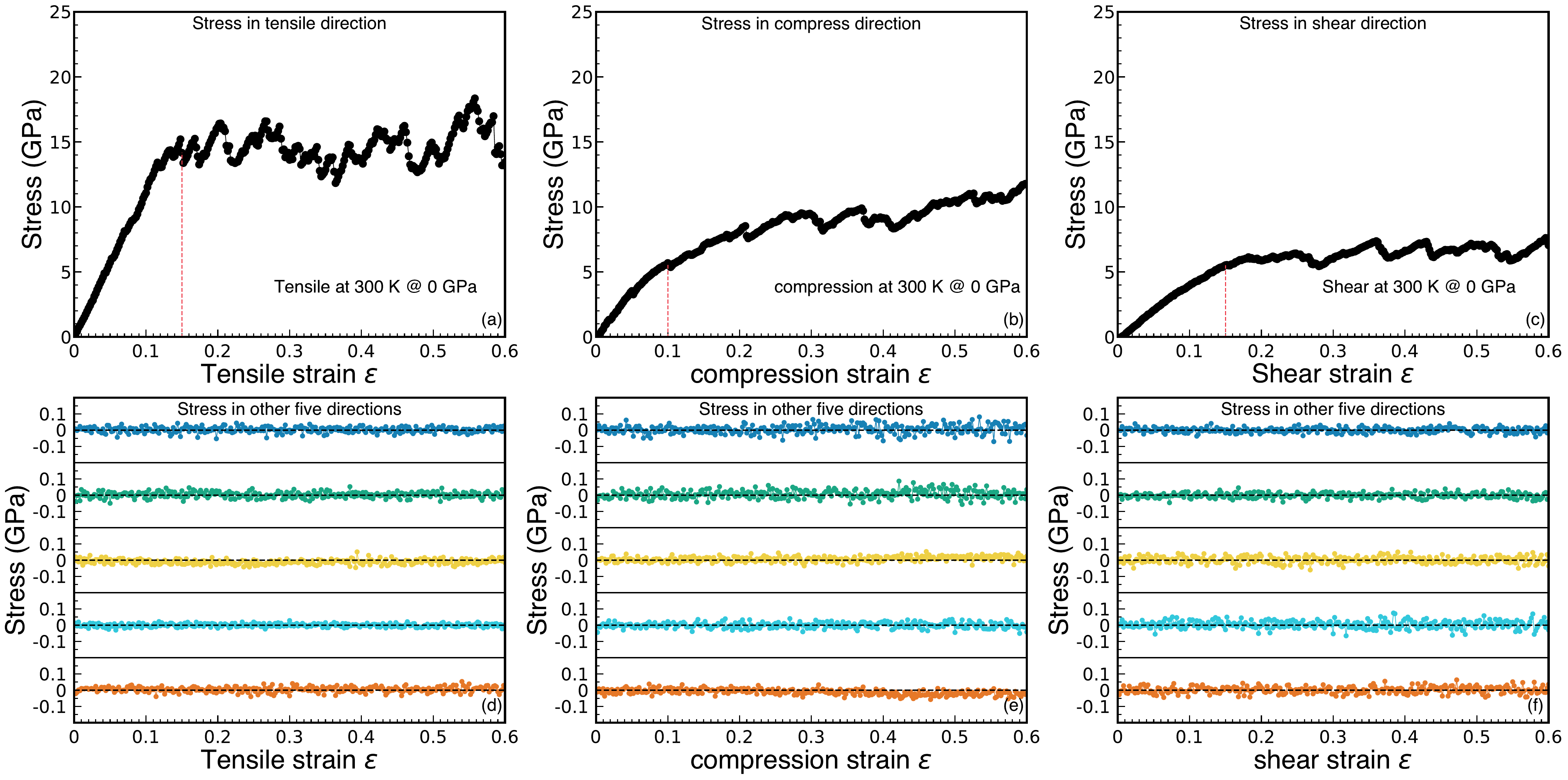}
	\caption{Upper panels, the averaged stress--strain curves of GC at 300~K under 0~GPa for tensile (a), compression (b), and shear (c) loading,  where the stresses are parallel to the directions of applied strains. The vertical dashed lines indicate the maximum strain at which the density and sp\textsuperscript{3} fraction remain within the typical range of GC, suggesting the upper limit of elastic behavior. Lower panels, the averaged stress--strain curves for other five stresses corresponding to tensile (d), compression (e), and shear (f) loading.
	Each curve represents an average over the three principal crystallographic directions and nearby 2000 data points.}
	\label{FIG:4}
\end{figure*}

\begin{table}[h]
	\centering
	\caption{The average density and percentage of sp\textsuperscript{2} bonds  
		of GC at different temperatures.}
	\setlength{\tabcolsep}{3pt}  
	\renewcommand{\arraystretch}{1.2} 
	\setlength{\arrayrulewidth}{1.2pt} 
	\label{tab1}
	\begin{tabular}{cccccc}
		\toprule[1.5pt] 
		\textbf{Property} & \textbf{300~K} & \textbf{1000~K} & \textbf{2000~K} & \textbf{2900~K} & \textbf{3000~K} \\
		\midrule[1.2pt] 
		\(\overline{\rho}\)~(g/cm\textsuperscript{3}) & 1.54 & 1.54  & 1.55 & 1.56  & 1.75 \\
		\(\overline{sp}\)\textsuperscript{2} (\%) & 94.8 & 94.8  & 95.4 & 95.3 & 96.7 \\
		\bottomrule[1.5pt] 
	\end{tabular}
\end{table}

In Tab.~\ref{tab1}, although the average density rises slightly with increasing temperature (up to 2900~K), it remains close to 1.5~g/cm\textsuperscript{3} , and percentage of sp\textsuperscript{2} bonds also stays near to 95\%. 
The sudden increase tendency in the average density and percentage of sp\textsuperscript{2} bonds at 2900$\sim$3000~K is consistence with the graphitization of GC observed under heat treatment at 3000~\textdegree C~\cite{Yamada_Nature_193_261_1962,Wang_Carbon_41_188_2003}. 
The low degree variation of the density and percentage of sp\textsuperscript{2} bonds demonstrate that the GC structure we prepared remains stable even at temperatures close to 3000~K, which aligns with previous expectations for GC structures~\cite{Yamada_Nature_193_261_1962,Yamaguchi_Carbon_1_47_1963,Cowlard_JMaterSci_2_507_1967,Wang_Carbon_41_188_2003}.

In addition, the tensile, compression, and shear deformation of GC are simulated using our developed CAIMD. We focus on the mechanical responses of GC to these external strains in ambient conditions at which GC has wide applications as wear-resistant composites~\cite{Myalski_Polymers_12_2264_2020} and corrosion protection systems~\cite{Sure_Carbon_67_643_2014} etc., with the simulated stress–strain curves given in Fig.~\ref{FIG:4}.
Each curve represents the average over three principal crystallographic directions to overcome the non-perfect isotropy of our GC model in Fig.~\ref{FIG:3}. Apart from the stresses parallel to the directions of applied strains, all the other five averaged stresses are less than 0.1~GPa, satisfying the criterion for the CAIMD described in the {\bf Methods of Calculations} section. For both tensile and shear deformation within a strain range of 0.15, the internal stresses increase linearly with the applied strains. We find that the density or sp\textsuperscript{2} fraction of the strained structure remain within the typical range of the starting GC. This suggests that these structures can elastically return to their original GC state upon strain releasing. In contrast, the compression strain induces more substantial structural changes, with the elastic region limited to strains below 0.1. After the elastic deformation, the internal stresses of GC either keep almost constant (for tensile or shear), or rise slowly (for compression), as the applied strains increase to above 0.6, which demonstrates the unexpectedly high plasticity of GC. This is very different from a previous classical MD calculation, where the internal stresses of GCs decrease quickly after reaching a peak linearly at a critical strain of around 0.2$\sim$0.3~\cite{Tomas_ApplPhysLett_112_251907_2018}. The discrepancy may arise from the fact that GCs with lower mass densities ($\leq$1.1 g/cm$^3$) were considered in the previous work~\cite{Tomas_ApplPhysLett_112_251907_2018}. It is worth noting that in classical MD, a pre-determined inter-atomic potential is used for the entire deformation processes under applied strains, while in our calculations the MLFF is updated by the first-priciples calculations during the strained deformation using the same supercell, which can describe more accurately the atomic bond braking and rebonding processes under large strains. Even though the internal stresses in Fig.~\ref{FIG:4} show structural stability under large strains up to 0.6, the GC structure actually changes markedly. The density and percentage of different sp bonds averaged over 1000 CAIMD steps and three principal crystallographic directions  at the strain of 0.6 under tensile, compression, and shear loading are listed in Tab.~\ref{tab-new}, which can be compared to those of equilibrium GC at 300~K in Tab.~\ref{tab1}. Surprisingly, the mass density of GC remains almost the same under a large tensile strain (0.6), which means a obvious shrinkage in the cross-section vertical to the tensile strain, as shown in Fig.~S3 of SM. The constant mass density maintains the high density of carbon bonds and unchanged tensile strength in GC. The increase of the sp bond percentage up to 10.6\% indicates that GC is becoming progressively unstable and will eventually break up under further tensile strains ($\epsilon$ > 0.6). It is interesting to see that densification of GC not only appears under compression, but also happens under pure shear, while the sp bonding distributions change little, which enhance the strength of GC in both cases. Such strain-stiffening effects are very useful for designing GC as protection coating materials. Many carbon materials, such as diamond and GC, are considered as brittle. But large elastic deformations of diamond predicted by first principles calculations have been confirmed experimentally in nano-sized diamond due to the significant reduction of defect densities in these nano-samples~\cite{Banerjee_Science_360_300_2018,Dang_Science_371_76_2021}. Given its amorphous nature of C-C bonds, we expect that the high plastic character of GC revealed by our CAIMD simulations can also be observed.

\begin{table}[h]
	\centering
	\caption{Average density and percentage of carbon atoms with sp, sp\textsuperscript{2}, and sp\textsuperscript{3} bonds at the strain of 0.6 under tensile, compression, and shear loadings.}
	\setlength{\tabcolsep}{3pt}  
	\renewcommand{\arraystretch}{1.2} 
	\setlength{\arrayrulewidth}{1.2pt} 
	\label{tab-new}
	\begin{tabular}{ccccc}
		\toprule[1.5pt] 
		\textbf{} & \textbf{\(\overline{\rho}\)~(g/cm\textsuperscript{3})} & \textbf{\(\overline{sp}\) (\%)} & \textbf{\(\overline{sp}\)\textsuperscript{2} (\%)} & \textbf{\(\overline{sp}\)\textsuperscript{3} (\%)}  \\
		\midrule[1.2pt] 
		Tensile & 1.56 & 10.6  & 87.6 & 1.8   \\
		Compression & 2.28 & 3.7  & 91.8 & 4.5  \\
		Shear & 2.11 & 3.8 & 93.2 & 3.0  \\ 
		\bottomrule[1.5pt] 
	\end{tabular}
\end{table}

\begin{figure*}
	\centering
	\includegraphics[width=0.9\textwidth]{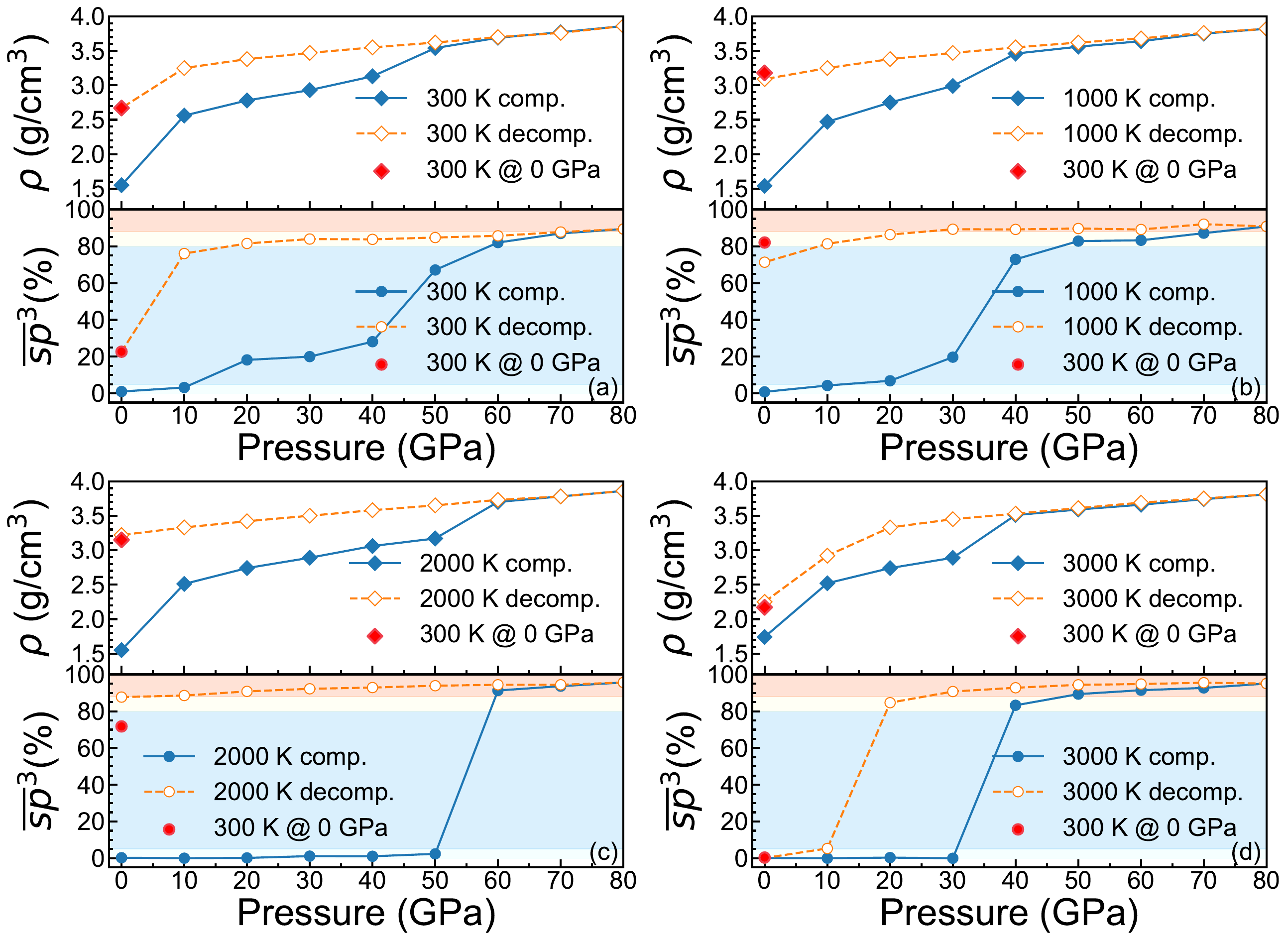}
	\caption{ The average density (upper panels) and the percentage of sp\textsuperscript{3} bonds (lower panels) as a function of pressure under various temperatures, (a) 300~K, (b) 1000~K, (c) 2000~K, and (d) 3000~K. Each data point represents the time-averaged value obtained from at least 50~ps of simulation after equilibration. Solid lines indicate stepwise compression, while dashed lines represent stepwise decompression. The red sample marks the final density or sp\textsuperscript{3}-bond percentage at 300~K and 0~GPa. The percentage of sp\textsuperscript{3} bonds is categorized into four distinct regions based on the bond percentage, each represented by a different color: 0--5\% (GC or graphite structure), 5--80\% (a-C structure), 8--88\% (ta-C structure), and 88--100\% (a-D structure).}
	\label{FIG:5}
\end{figure*}

\begin{figure*}
	\centering
	\includegraphics[width=0.9\textwidth]{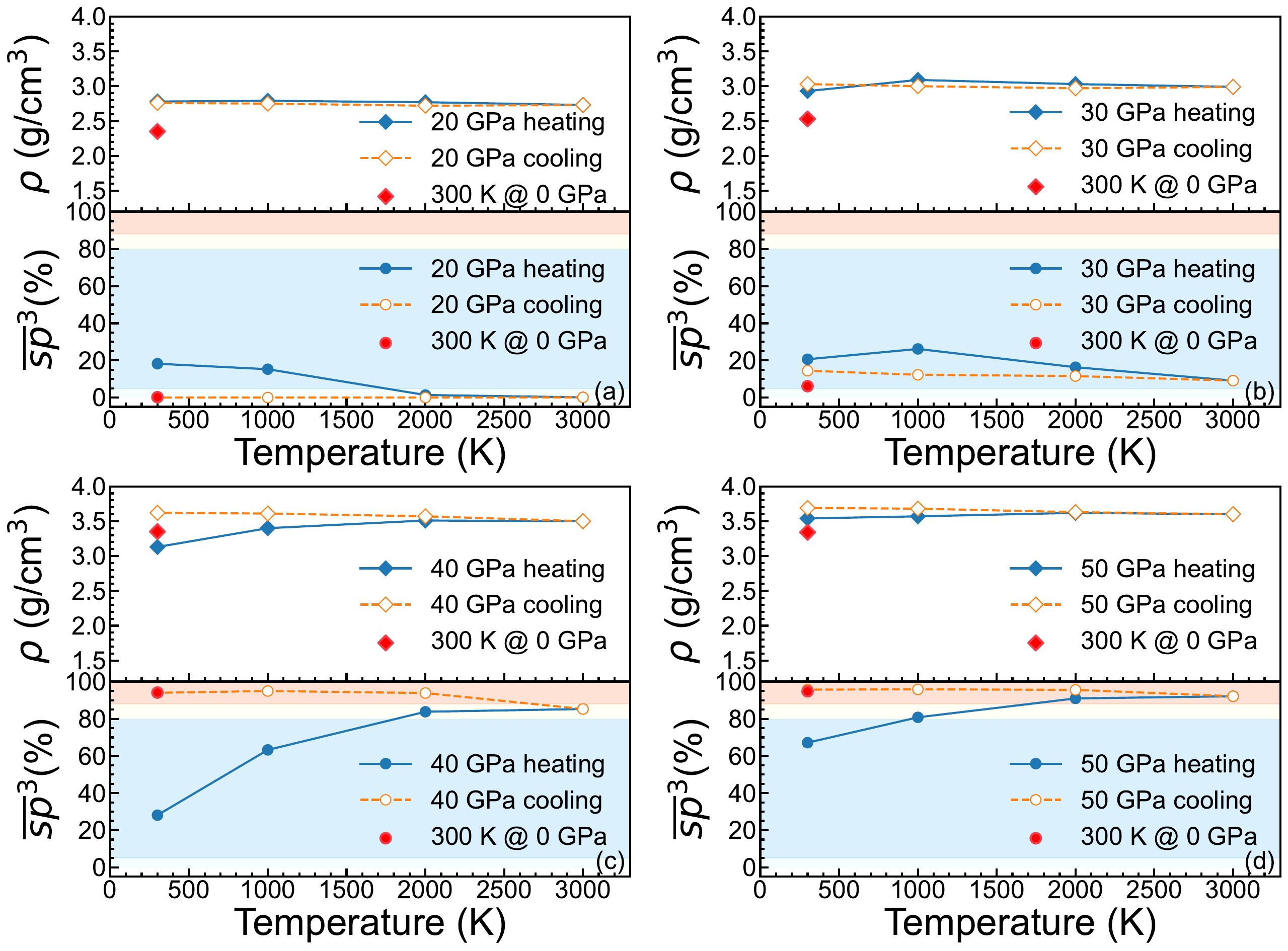}
	\caption{ The average density (upper panels) and the percentage of sp\textsuperscript{3} bonds (lower panels) as a function of temperature under various pressures, (a) 20~GPa, (b) 30~GPa, (c) 40~GPa, and (d) 50~GPa. Each data point represents the time-averaged value obtained from at least 50~ps of simulation after equilibration. Solid lines indicate stepwise heating, dashed lines represent stepwise cooling. The red sample marks the final density or sp\textsuperscript{3}-bond percentage at 300~K and 0~GPa. The percentage of sp\textsuperscript{3} bonds is categorized into four distinct regions with different colors based on the bond percentage, similar to that in Fig.~\ref{FIG:5}}.
	\label{FIG:6}
\end{figure*}

Now we start to investigate the transformation of GC into ultrahard materials, such as a-D, by applying a series of pressure, temperature, uniaxial compression, and severe rotational shear strains to simulate the structural evolution processes of GC. We begin with considering only temperature and hydrostatic pressure, which have been extensively studied before experimentally and theoretically~\cite{Yamada_Nature_193_261_1962,Wang_Carbon_41_188_2003,Lin_PhysRevLett_107_175504_2011,Yao_ApplPhysLett_104_021916_2014,Yao_ApplPhysLett_111_101901_2017,Shiell_PhysRevLett_120_215701_2018,Zeng_PhysRevMater_3_033608_2019,Shiell_PhysRevB_99_024114_2019,Zeng_PhysRevB_109_214113_2024,Huang_Carbon_219_118763_2024}. As a complement, we examine the effects of different orders when high pressure and temperature are exerted on GC, using the original VASP on-the-fly ML-augmented \textit{NpT} AIMD. In Fig.~\ref{FIG:5}, the average percentage of sp\textsuperscript{3} bonds and mass density are plotted, for  the equilibrium configurations of GC pre-heated at 300, 1000, 2000, and 3000~K, as functions of pressure from 0 to 80~GPa (solid curves) and then completely decompressed (dashed curves) followed by cooling down to 300~K (red symbols). In Fig.~\ref{FIG:6}, the same results are presented, for the equilibrium GC samples pre-pressured at 20, 30, 40, and 50~GPa, as functions of temperature from 300 to 3000~K (solid curves) and then cooled down to 300~K (dashed curves) followed by decompression to 0~GPa (red symbols).
At each temperature and pressure point, a simulation of at least 100~ps is conducted, and all the data is obtained by averaging over the last 50~ps after reaching equilibrium.
We classify the transformed structures of GC under pressure and temperature, or later involving various stresses, into four distinct categories based on the percentage of sp\textsuperscript{3} bonds~\cite{McKenzie_RepProgPhys_59_1611_1996,Fallon_PhysRevB_48_4777_1993,Pharr_ApplPhysLett_68_779_1996,Tan_JPhysChemC_124_5489_2020,Klein_Carbon_107_536_2016,Seok_SciRep_8_13521_2018,Zhao_JApplPhys_135_065304_2024,Lin_PhysRevLett_107_175504_2011,Shiell_SciRep_6_37232_2016,Zeng_NatCommun_8_322_2017,McCulloch_Small_16_2004695_2020,Huang_Carbon_219_118763_2024}: (i) The sp\textsuperscript{3} percentage is below 5\%, 
so the structure either remains as GC or is mostly graphitized; (ii) The sp\textsuperscript{3} percentage is between 5\% and 80\%, where the resulting structure exhibits much lower hardness compared to diamond, yet maintains substantial ductility, commonly known as a-C; (iii) The sp\textsuperscript{3} percentage ranges between 80\% and 88\%, where the material demonstrates hardness approaching that of diamond (up to approximately 80~GPa), and is named ta-C; (iv) The sp\textsuperscript{3} percentage exceeds 88\%, so the hardness further increases, leading to the formation of a-D or nanocrystalline structures of cubic or hexagonal diamond. Different colors are used in our figures to distinguish the four regimes mentioned above.

We first discuss the compression processes with different preheated temperatures in Fig.~\ref{FIG:5}. To illustrate the critical behavior at 3000~K, the results obtained from identical simulations performed at 2900~K is also present in Fig.~S4. It can be observed that at low temperatures (such as 300 and 1000~K), the sp\textsuperscript{3} percentage increases gradually with pressure and then quickly goes up when the pressure is higher than 40~GPa (at 300~K) or 30~GPa (at 1000~K), while at high temperatures (such as 2000~K, 2900~K, and 3000~K),  the sp\textsuperscript{3} percentage keeps low ($<$5\%) and constant, and then rises suddenly when the pressure is above a threshold.
The stark contrast between low and high temperatures is primarily attributed to the graphitization of GC at elevated temperatures. Experimentally, it is found that at low temperature, the structural transformation of GC under increasing pressure follows the sequence: GC \(\rightarrow\) sp\textsuperscript{2}-sp\textsuperscript{3} carbon \(\rightarrow\) sp\textsuperscript{2}-sp\textsuperscript{3} + graphite (or diamond), eventually transitioning into a-D. However, when the temperature exceeds 1500~\textdegree C, the GC bypasses the sp\textsuperscript{2}-sp\textsuperscript{3} carbon phase and directly undergoes graphitization~\cite{Hu_JMateriomics_7_177_2021}. This is fully inline with our results. The gradual increase of sp\textsuperscript{3} percentage in Fig.~\ref{FIG:5}(a,b) indicates the formation of sp\textsuperscript{2}-sp\textsuperscript{3} carbon at low temperature which eventually transforms into a-D as pressure goes up. The low constant sp\textsuperscript{3} percentage and growing density in Fig.~\ref{FIG:5}(c,d) and Fig.~S4 reveals graphitization process at high temperature which gives rise to a sudden increase in sp\textsuperscript{3} percentage due to the crumple of nearing graphite layers under high pressure and thermal vibrations. The critical pressure (50, 40, 30~GPa) for the sudden increase of sp\textsuperscript{3} percentage gets lower as the temperature (2000, 2900, 3000~K) or thermal vibrations elevates.

It has been well demonstrated that the high sp\textsuperscript{3} percentage in GC induced by cold compression at room temperature is not stable after decompression~\cite{Lin_PhysRevLett_107_175504_2011,Yao_ApplPhysLett_104_021916_2014,Yao_ApplPhysLett_111_101901_2017,Shiell_PhysRevLett_120_215701_2018,Zeng_PhysRevMater_3_033608_2019,Zeng_PhysRevB_109_214113_2024,Huang_Carbon_219_118763_2024}. An classical MD calculation with a supercell of 32,768 atoms shows that the sp\textsuperscript{3} percentage and density of GC are about 22\% and 2.4 g/cm\textsuperscript{3}after decompression from 80 GPa at 300~K~\cite{Huang_Carbon_219_118763_2024}, close to our results (22.7\% and 2.67 g/cm\textsuperscript{3}) in Fig.~\ref{FIG:5}(a). But their predicted sp\textsuperscript{3} percentage of 34\% at a pressure of 80~GPa~\cite{Huang_Carbon_219_118763_2024} is too low compared to that (75\%) detected in an experiment~\cite{Zeng_PhysRevB_109_214113_2024}, which reports a hysteresis loop of sp\textsuperscript{3} percentage curve overall agrees with our result in Fig.~\ref{FIG:5}(a), only slight lower by 10$\sim$15\%. The calculated decompression sp\textsuperscript{3} percentage and high density in Fig.~\ref{FIG:5}(b, c) and Fig.~S4 also are consistent with a recent experiment~\cite{Zeng_PhysRevB_109_214113_2024} that high temperature annealing can preserve the major sp\textsuperscript{3} percentage generated by high pressure after decompression even at 2900~K. However, as shown in Fig.~\ref{FIG:5}(d), with further increase of temperature to 3000~K, the finial a-D structure at 80~GPa dissolves during decompression as it is close to the graphitization temperature (3000~$^{\circ}$C) of GC at ambient pressure~\cite{Yamada_Nature_193_261_1962}, evidenced by a graphite density of 2.2 g/cm$^3$. The effect of cooling on the resulted carbon structures after decompression is minor, as revealed in Fig.~\ref{FIG:5} and Fig.~S4.

We turn to discuss a different effect on GC by reversing the order of pressuring and heating.  Zeng et al. reported synthesizing an a-D sample by laser heating GC to 1800~K after first compressing them to 50~GPa in DAC at room temperature~\cite{Zeng_NatCommun_8_322_2017}. 
In Fig.~\ref{FIG:6}, we present heating simulations on GC, which is first pressured to 20, 30, 40, and 50~GPa at room temperature (300~K) and then gradually increased to 3000~K, where a similar structural transformation to a-D~\cite{Zeng_NatCommun_8_322_2017} is observed in Fig.\ref{FIG:6}(d). At each temperature point, we conducted at least 100~ps of simulation and averaged the data over the final 50~ps after equilibration to determine the average sp\textsuperscript{3}-bond percentage under each temperature and pressure condition. As before, solid lines represent the heating process, while dashed lines show the cooling. The structure is also categorized into four regions, similar to the classification in Fig.~\ref{FIG:5}. It can be observed that with increasing temperature, the sp\textsuperscript{3} percentage exhibits distinctly different trends for structures initially at pressures \(\leq\) 30~GPa and \(\geq\) 40~GPa, which makes the beginning sp\textsuperscript{3} percentage of the sample at 300~K below or above 30\%. We believe this is a threshold above which sp\textsuperscript{3} percentages of the carbon structure can increase with rising heat-treatment temperature and finally transform into a-D at sufficiently high annealing temperatures (\(>\)2000~K). Also, when the pressure is higher than 40~GPa, the order of pressuring and heating matters critically on the evolution of GC structures. For instance, if GC is pre-heated to 2000$\sim$2900~K, a pressure of 40~GPa produce no sp\textsuperscript{3} bonds (see Fig.~\ref{FIG:5}(c) and Fig.~S4). However, if GC is pre-pressured to 40~GPa, heating to 2000$\sim$3000~K results in ta-C or a-D structures (see Fig.~\ref{FIG:6}(c)). Similarly, if the resulted a-D is decompressed from 40~GPa followed by cooling from 3000~K to 300~K, a graphite is obtained, as shown in Fig.~\ref{FIG:5}(d). But if the a-D is cooled down from 3000~K to 300~K followed by decompression from 40~GPa, it remains in the a-D structure, as in Fig.~\ref{FIG:6}(c).  

As the carbon structures at different temperatures pre-pressured at 30 and 40~GPa in Fig.~\ref{FIG:6}(b, c) will be used below to simulate uniaxial compression and rotational shear deformations of GC, the reliability of their critically different structure evolution under a pressure of 30 or 40~GPa is tested by three additional heating simulations at these pressures, as presented in Fig.~S5 of SM. The structural evolution in Fig.~\ref{FIG:6}(b, c) and Fig.~S5 are highly consistent,
which verifies that the critical pressure threshold for the sp\textsuperscript{2}-to-sp\textsuperscript{3} transition in GC lies between 30 and 40~GPa, in agreement with earlier reported pressures required for initiating the transformation from sp\textsuperscript{2} to sp\textsuperscript{3} hybridization in GC (40~GPa in Ref.~\cite{Lin_PhysRevLett_107_175504_2011} and 33~GPa in Ref.~\cite{Yao_ApplPhysLett_104_021916_2014}). Based on our results, we propose that compressing GC initially to 40~GPa (10~GPa lower than the pressure reported by Zeng \textit{et al.}~\cite{Zeng_NatCommun_8_322_2017}) and subsequently heating it to 3000~K (achievable through laser heating) is sufficient to induce a transformation to a-D.

\begin{figure*}
	\centering
	\includegraphics[width=0.9\textwidth]{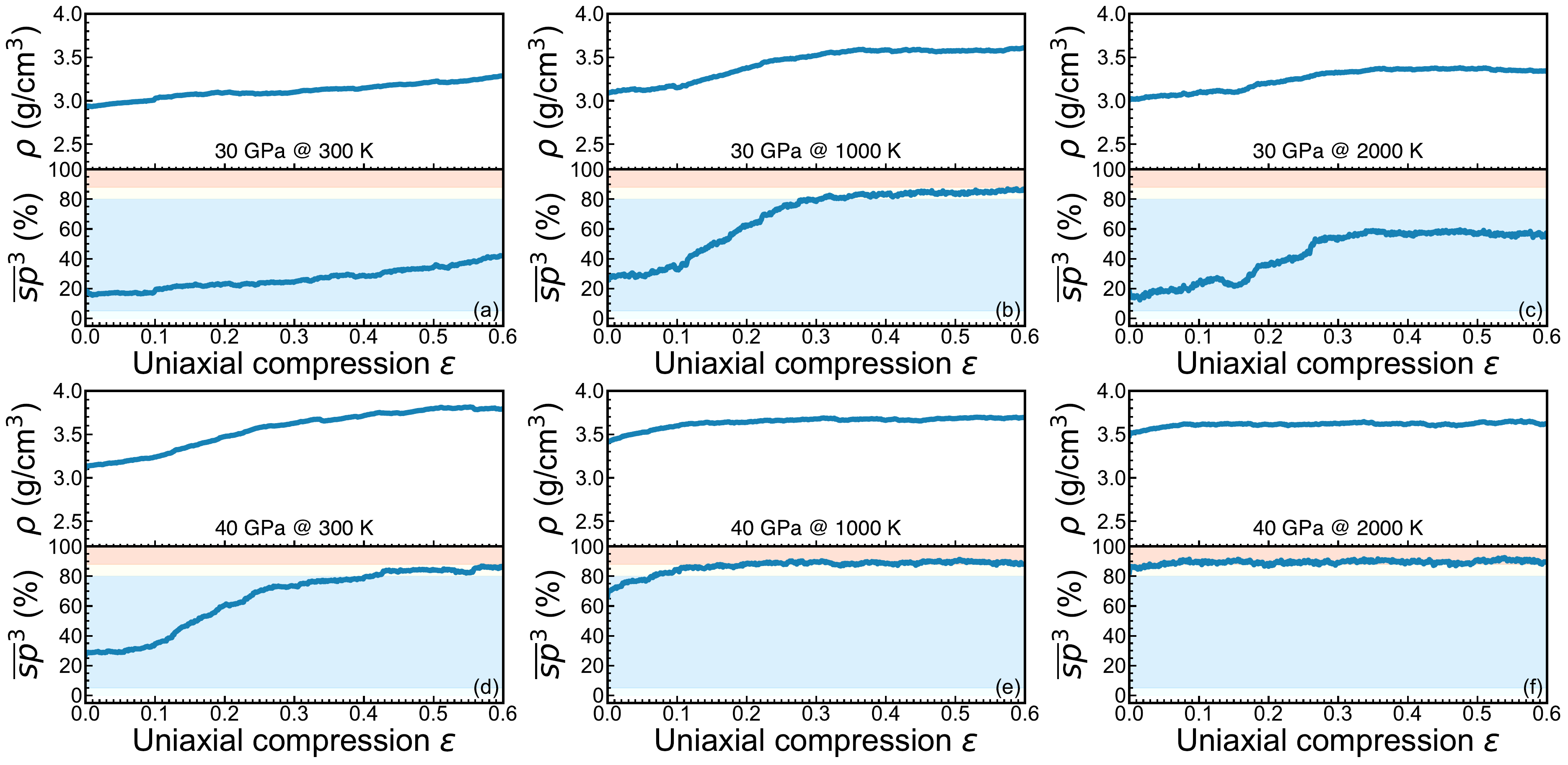}
	\caption{The percentage of sp\textsuperscript{3} bonds and mass density as a function of uniaxial compression under a hydrostatic pressure of 30~GPa (upper panels) and 40~GPa (lower panels) at different temperatures: (a,d) 300~K, (b,e) 1000~K, and (c,f) 2000~K. At each temperature, uniaxial compression is applied along three distinct crystalline directions, \(\left[100\right]\), \(\left[010\right]\), and \(\left[001\right]\). Each data point was sampled at 1~ps (1000 AIMD stpes) intervals and averaged over three crystallographic directions. From bottom to top, the material is categorized into four distinct regions with different colors based on the bond percentage, similar to that in Fig.~\ref{FIG:5}}.
	\label{FIG:7}
\end{figure*}

Our calculations so far have demonstrated that the on-the-fly ML-augmented AIMD simulations we employed exhibit excellent accuracy, reproducing experimental results in many cases. Notably, we showed that by raising the temperature to 3000~K, the pressure required to induce the GC to a-D transformation could be reduced to 40~GPa. However, previous studies have repeatedly pointed out that non-hydrostatic compression and shear stress play a crucial role in the structural transformation of GC~\cite{McCulloch_Small_16_2004695_2020,Huang_Carbon_219_118763_2024,Yao_ApplPhysLett_104_021916_2014}. The on-the-fly ML-augmented CAIMD simulations we implemented by modifying the original VASP code (see {\bf Methods of Calculations}) provide a powerful means to verify these conjectures. In the following, we will separately examine the possibilities by non-hydrostatic compression and severe rotational shear strains to reduce the required pressure and temperature for GC transformation to hardened structures.

Firstly, for non-hydrostatic compression, due to the extremely low hardness of GC, directly applying uniaxial compression in a DAC leads to rapid structural collapse (see Fig.~\ref{FIG:4}(b)). A straightforward solution is to use a GC sample with its diameter close to the thickness of gasket filled with pressure medium, so under continuous compressing the stress at the contacting point between the DAC and GC sample is different from the confining pressure in the gasket as the uniaxial strain applies~\cite{Lin_PhysRevLett_107_175504_2011,Yao_ApplPhysLett_104_021916_2014,Yao_ApplPhysLett_111_101901_2017}. The large differences between the compressing stresses and confining pressures indicate the superhardness of uniaxially compressed GC structure. But to best of our knowledge, none accurate theoretical simulation exists to explain the largest compressing stress reachable for a given confining pressure. Another way to uniaxially compress GC is to fill the gasket of DAC with GC and without pressure medium, the process of which has been simulated by classical MD where the lateral cross section of the GC sample is fixed during compression~\cite{Shiell_PhysRevLett_120_215701_2018,Huang_Carbon_219_118763_2024}. Different results are reported in these classical MD simulations, where in one case the sp\textsuperscript{3} percentage reaches over 60\% under the largest compression stress of 55~GPa~\cite{Shiell_PhysRevLett_120_215701_2018}, while in another case, it is only 20\% (or 40\%) under a stress of 55~GPa (or 80~GPa)~\cite{Huang_Carbon_219_118763_2024}. 
We focus on the non-hydrostatic compression with finite confining pressures~\cite{Lin_PhysRevLett_107_175504_2011,Yao_ApplPhysLett_104_021916_2014,Yao_ApplPhysLett_111_101901_2017}, which can be studied using the CAIMD algorithm we developed and find that a noticeable increase in the percentage of sp\textsuperscript{3} bonds only occurs when uniaxial compression is applied under a confining pressure of at least 30~GPa.

\begin{figure*}
	\centering
	\includegraphics[width=0.9\textwidth]{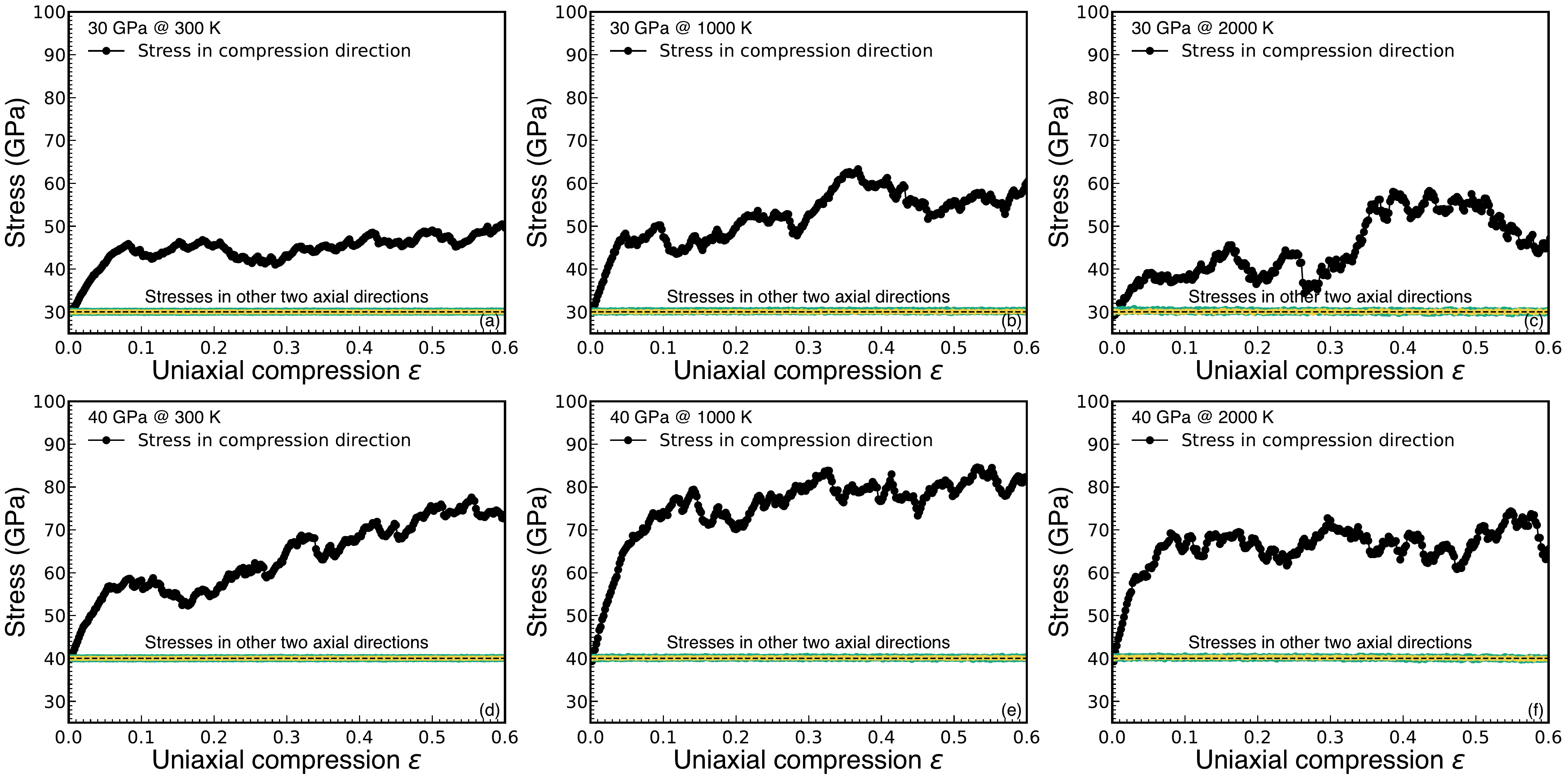}
	\caption{ The average stresses, in directions parallel and perpendicular to the uniaxial compression direction, as a function of the compression strain under a hydrostatic pressure of 30~GPa (upper panels) and 40~GPa (lower panels) at different temperatures: (a,d) 300~K, (b,e) 1000~K, and (c,f) 2000~K. For each uniaxial strain value, the stress is averaged over 2000 data points in its vicinity. The compression strains are applied to three distinct crystallographic orientations, with the final curves represent an average over three orientation directions.} 
\label{FIG:8}
\end{figure*}

Fig.~\ref{FIG:7} shows the percentage of sp\textsuperscript{3} bonds and mass density as a function of uniaxial compression 
under a confining pressure of 30 and 40~GPa at 300, 1000, and 2000~K, 
where compression consistently promotes the formation of sp\textsuperscript{3} bonds over all temperatures. 
However, the effect of temperature on the increase of sp\textsuperscript{3} percentage with uniaxial compression strains is not monotonic, which is much faster at 1000~K than that at 300~K under a confining pressure of 30~GPa, but becomes obviously slower at 2000~K due to the much reduced graphitization temperature (1500~\textdegree C) of GC under pressures~\cite{Hu_JMateriomics_7_177_2021}. For a confining pressure of 40~GPa, the increase of sp\textsuperscript{3} percentage with uniaxial compression strains become increasingly faster when temperature is higher, with the final structure all evolving into ta-C or a-D. 
Comparing the result in Fig.~\ref{FIG:7}(b) (30~GPa at 1000~K) with the hydrostatic case in Fig.~\ref{FIG:6}(c), where 40~GPa is needed to induce ta-C at a higher temperature of 2000~K, the required pressure is reduced by approximately 10~GPa, and the transition occurs at a more experimentally accessible temperature of 1000~K.
As shown in Fig.~\ref{FIG:7}(d-f), when the pressure is over 40~GPa, exceeding the threshold at which sp\textsuperscript{3} bonds begins to massively appear in GC, a-D structures can be formed via uniaxial compression even at 300~K. This further confirms the existence of a critical pressure threshold in the range of 30--40~GPa. Once this threshold is surpassed, GC can easily transforms into ta-C or a-D, either through high temperature or non-hydrostatic compression. 
In order to further understand why non-hydrostatic compression is effective in promoting the formation of a-D, we analyze the compressive stress evolution during compression in Fig.~\ref{FIG:8}. The results reveal that for the confining pressure fixed at 30 or 40~GPa, the stress along the compressed direction significantly exceeds 60~GPa in Fig.~\ref{FIG:8}(b,d-f), matching the pressure levels typically required to achieve ta-C or a-D in experiments. Therefore, it is not surprising that such uniaxial compression can effectively induce the formation of ta-C or a-D structures (see Fig.~\ref{FIG:7}(b,d-f)).

The stress evolution averaged over three compression directions and nearby 2000 CAIMD steps at 300, 1000, and 2000~K under a confining pressure of 30~GPa or 40~GPa is presented in Fig.~\ref{FIG:8}. For each case, we also show the average stresses in the two directions perpendicular to the compression axis. It can be seen that the stresses in these perpendicular directions match the confining pressure well, while the three shear stresses are averaged to zeros (not shown in Fig.~\ref{FIG:8}), all with errors below 0.1 GPa, indicating that our algorithm effectively satisfy the \textit{NpT} ensemble statistical behavior in unconstrained directions. In contrast, the evolution of stresses along the compression direction differs significantly for the two confining pressures due to the distinct resulting sp\textsuperscript{3} fractions in GC (see Fig.~\ref{FIG:8}(a-c) and Fig.~\ref{FIG:8}(d-f)). At 300~K, the stresses under 30 and 40~GPa confining pressures both rise quickly at the beginning of compression, then the former becomes almost constant at 50~GPa, while the latter increases continuously to 75~GPa, as shown in Fig.~\ref{FIG:8}(a,d), due to the different sp\textsuperscript{3} fractions induced by compression strains (see Fig.~\ref{FIG:7}(a,d)). The calculated compression stress (or differences between the compression stress and confining pressure) agree well with previous experimental results, which is 40~GPa at a 30~GPa confining pressure and 70~GPa at a 40~GPa confining pressure~\cite{Lin_PhysRevLett_107_175504_2011,Yao_ApplPhysLett_104_021916_2014}. The effects of high temperatures on the compression stresses are also interesting. At 1000~K, the compression stresses are all enhanced comparing to those at 300~K, as shown in Fig.~\ref{FIG:8}(b,e), because high thermal energies promote sp\textsuperscript{3}-bond formation (see Fig.~\ref{FIG:7}(b,e)). With further increase of temperature to 2000~K, the compression stresses are weakened, as shown Fig.~\ref{FIG:8}(c,f), for two reasons: (i) if the sp\textsuperscript{3} fraction is high (see Fig.~\ref{FIG:7}(f)), its compression strength weakens at higher temperature as in diamond~\cite{Wen_Carbon_155_361_2019}; (ii) if the confining pressure is below 30~GPa, the reduced graphitization temperature (1500~\textdegree C) of GC under pressures~\cite{Hu_JMateriomics_7_177_2021} results in a low sp\textsuperscript{3} fraction (see Fig.~\ref{FIG:7}(c)) which further decreases the compression stresses.

\begin{figure*}
	\centering
	\includegraphics[width=0.9\textwidth]{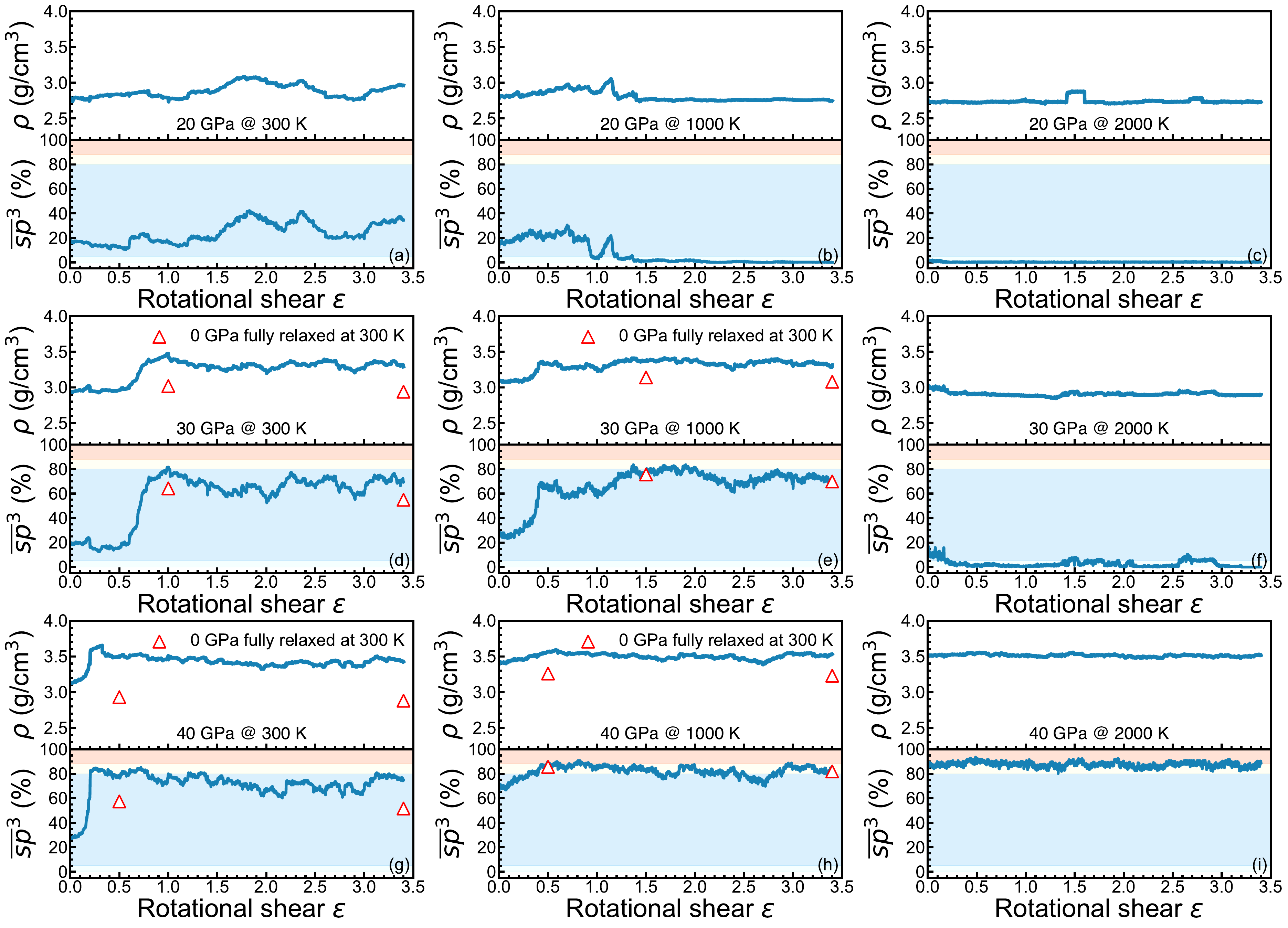}
	\caption{The percentage of sp\textsuperscript{3} bonds and mass density as a function of rotational shear strain under a hydrostatic pressure of 20~GPa (upper panels), 30~GPa (middle panels) and 40~GPa (lower panels) at different temperatures: (a-c) 300~K, (d-f) 1000~K, and (g-i) 2000~K. 
	Data points were recorded at intervals of 1~ps (every 1000 AIMD steps). From bottom to top, the material is categorized into four distinct regions with different colors based on the bond percentage, similar to that in Fig.~\ref{FIG:5}. The triangle symbols represent the calculated sp\textsuperscript{3} percentage and mass density for GC structures under selected rotational shear strains after cooling to 300~K by a 100~ps AIMD simulation followed by another 100~ps AIMD relaxation under 0~GPa (decompression).}
\label{FIG:9}
\end{figure*}

Finally, we apply rotational shears to pre-pressured GC, imitating the RDAC technique used in experiments, as sketched in Fig.~S1 of SM. Specifically, a shear strain of 0.2 is applied in each shear direction, with the material rotated by 22.5\textdegree~in each step for a total of 16 steps. In order to retain shear deformation in the final structure, we first imposed a shear strain of 0.2 prior to initiating the rotational shear process. This initial shear is reflected in the 0--0.2 region of the shear strain axis in Fig.~\ref{FIG:9}.
In total, a rotational shear strain of 3.4 is applied. Considering a strain rate of 0.001~ps$^{-1}$ and a timestep of 1~fs, the entire simulation required 3400~ps, corresponding to 3,400,000 AIMD steps. With the assistance of on-the-fly ML-augmented CAIMD  we developed, the simulation for one rotation circle is completed within one week or less while maintaining the accuracy of conventional AIMD.

In Fig.~\ref{FIG:9}, we systematically investigate how the percentage of sp\textsuperscript{3} bonds evolves with rotational shear strains under different isotropic pressures (20~GPa, 30~GPa, and 40~GPa) at temperatures of 300, 1000, and 2000~K. Data are recorded at intervals of 1~ps (every 1000 steps). We hope, by properly combining the effects of pressure, temperature and severe rotational shear, to find new synthesizing routes from GC to quenchable amorphous hard carbon allotropes with low pressure and temperature. As shown in Fig.~\ref{FIG:9}(a-c), if the pressure is lower than 20~GPa, rotational shear strains at room temperature are insufficient to induce a sharp increase in the percentage of sp\textsuperscript{3} bonds. With increasing temperature, the sp\textsuperscript{3} percentage even decreases to nearly zero, indicting that in this case, rotational shear strains assist to reduce the graphitization temperature in GC to below 1000~K. These results reveal that although rotational shear strains play an important role in promoting the transformation of GC into a-D, a relatively high pressure remains indispensable.

Under a pressure of 30~GPa, which is below the critical pressure (in between 30--40~GPa) for the massive formation of sp\textsuperscript{3} bonds in GC as shown in Fig.~\ref{FIG:5} and Fig.~\ref{FIG:6}, the effect of rotational shear strains differ significantly between low temperatures ($\le$1000~K in Fig.~\ref{FIG:9}(d,e)) and high temperatures ($\ge$2000~K in Fig.~\ref{FIG:9}(f)). 
The percentage of sp\textsuperscript{3} bonds increases sharply with rotational shear strains at low temperatures, especially in the 1000~K case even reaching the ta-C region (Fig.~\ref{FIG:9}(e) for shear strains $\epsilon$ $\approx$ 1.4$\sim$2.0). 
This implies that when the pressure is slightly below the critical pressure (in between 30--40~GPa), the non-uniform stress distribution induced by rotational shear deformation at low temperatures can cause the local pressure 
to exceed the threshold, promoting the generation of sp\textsuperscript{3} bonds and leading to a significantly high overall  percentage of sp\textsuperscript{3} bonds in the material. This may also explain why shear forces facilitate the formation of nanocrystalline diamond from amorphous materials under low pressure and temperature environments~\cite{McCulloch_Small_16_2004695_2020,Wong_Carbon_142_475_2019,Gao_Carbon_146_364_2019}. However, as the temperature increases to 2000~K, the sp\textsuperscript{3} percentage in GC decreases from start due to graphitization of GC,  which is accelerated by the applied rotational shear strains, resulting in an sp\textsuperscript{3} fraction close to zero. As the pressure further increases to 40~GPa, the effect of rotational shear strains on the structural evolution of GC remains basically the same as that under a pressure of 30~GPa, only with a quicker rise of sp\textsuperscript{3} fraction at earlier rotational shear strains at temperatures equal or below 1000~K, as shown in Fig.~\ref{FIG:9}(g,h), and the disappearance of rotational shear assisted graphitization of GC at 2000~K, as depicted in Fig.~\ref{FIG:9}(i).

\begin{figure*}
	\centering
	\includegraphics[width=0.9\textwidth]{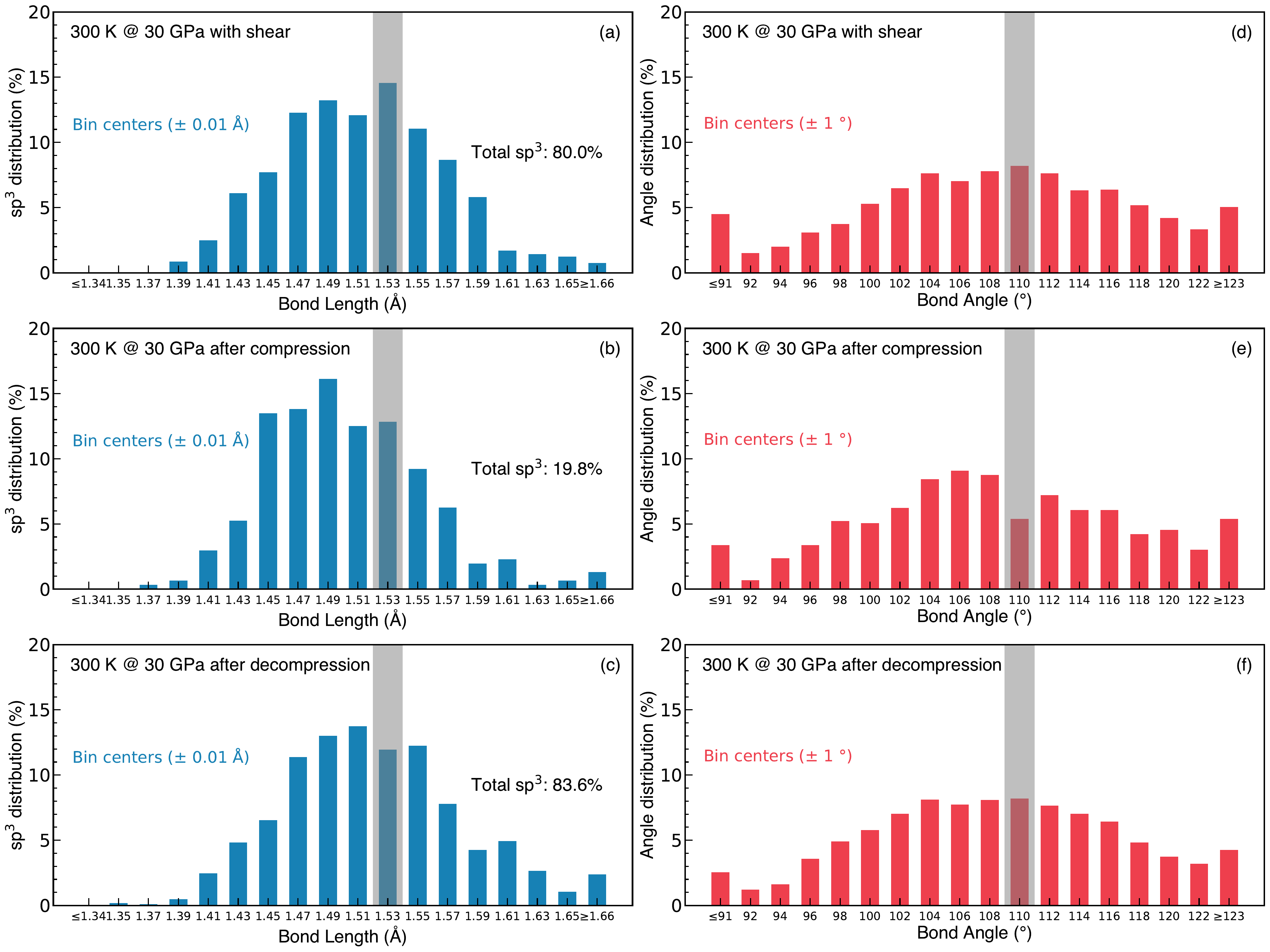}
	\caption{(a)--(c) show the bond length distributions of sp\textsuperscript{3} bonds in GC at room temperature (a) under 30~GPa with a rotational shear $\epsilon$=1.0, (b) after compression from equilibrium to 30~GPa, and (c) after decompression from 80~GPa to 30~GPa, respectively, while (d)--(f) show the corresponding bond angle distributions under the same conditions. Gray bins indicate the reference distributions of sp\textsuperscript{3} bonds in cubic diamond at the same temperature and pressure.}
	\label{FIG:10}
\end{figure*}

What is more surprising is that the high sp\textsuperscript{3}-bonding carbon structures formed under rotational shear strains with proper pressure and temperature are quenchable to normal ambient conditions. In Fig.~\ref{FIG:9}(d,e,g,h), we plot the calculated sp\textsuperscript{3} percentage and mass density (triangle symbols) for GC structures under selected rotational shear strains after cooling to 300~K by a 100~ps AIMD simulation (if the temperature is higher than 300~K) followed by another 100~ps AIMD relaxation under 0~GPa (decompression). In each case, we choose the rotational shear strain at which the sp\textsuperscript{3} percentage is relatively or at the end of rotational shear. The results (triangle symbols) in Fig.~\ref{FIG:9}(d,e,g,h) show that all the quenched samples have sp\textsuperscript{3} percentages at least larger than 55\%, with some of them above 80\%. It is interesting to note that at 300~K under a pressure of 30~GPa, after a rotational shear strain of 1.0, the quenched sample maintains a sp\textsuperscript{3} percentage of 64\%, which to our knowledge is the easiest condition to synthesize amorphous carbon allotropes from GC with the quenchable sp\textsuperscript{3} percentage higher than 50\%. Experimentally, quenchable a-C with sp\textsuperscript{3} percentage of 50\% is obtained by applying a pressure of 58~GPa to GC followed by annealing at 728~K~\cite{Zeng_PhysRevB_109_214113_2024}, and quenchable a-D by a pressure of 50~GPa followed by laser heating to 1800~K~\cite{Zeng_NatCommun_8_322_2017}. Experimentally, all the hardened carbon samples from GC by cold compression, either hydrostatic or non-hydrostatic, are unstable after decompression to ambient condition~\cite{Lin_PhysRevLett_107_175504_2011,Yao_ApplPhysLett_111_101901_2017,Shiell_PhysRevLett_120_215701_2018,Zeng_PhysRevB_109_214113_2024}, 
consistent with our calculations in Fig.~\ref{FIG:5}(a), where for the structure compressed from GC under a high pressure of 80~GPa at 300~K, its sp\textsuperscript{3} fraction drops significantly from the initial 89\% to only 23\% upon pressure release. 

To understand the effect of rotational shear strains, Fig.~\ref{FIG:10} presents a comparison of the bond length and bond angle distributions of the sp\textsuperscript{3} atoms in the carbon structures at 300~K, corresponding to three structures obtained under distinct conditions:  
Fig.~\ref{FIG:10}(a,d) shows the structure subjected to a rotational shear strain of $\epsilon = 1.0$ at 30~GPa (from Fig.~\ref{FIG:9}(d));  
Fig.~\ref{FIG:10}(b,e) corresponds to a structure directly compressed to 30~GPa under hydrostatic conditions (from Fig.~\ref{FIG:5}(a));  
and Fig.~\ref{FIG:10}(c,f) depicts the structure decompressed from 80~GPa to 30~GPa (also from Fig.~\ref{FIG:5}(a)).
All the bond angle distributions are nearly identical, showing no significant effect from rotational shear strains. However, the bond length distributions exhibit clear differences. 
In the rotationally sheared structure (Fig.~\ref{FIG:10}~(a)), the bond lengths are centered around 1.51~\AA, similar to the length of sp\textsuperscript{3} bonds in cubic diamond under the same pressure. These are stable sp\textsuperscript{3} bonds with low energies,
which tend to remain as sp\textsuperscript{3} bonds after the removal of external pressure and rotational shear strains. In contrast, as depicted in Fig.~\ref{FIG:10}~(b,c), the sp\textsuperscript{3}-bond lengths in the directly compressed and decompressed structures under 30~GPa without rotational shearing, are mostly shorter bonds comparing to that of sp\textsuperscript{3} bonds in cubic diamond under the same pressure. These sp\textsuperscript{3} bonds are deformed with high energies, which easily break into sp\textsuperscript{2} bonds after pressure releasing. The same mechanism was exploited recently to explain the synthesizing of a-C with a quenchable sp\textsuperscript{3} fraction of 50\% under a pressure of 58~GPa followed by high temperature (728~K) annealing which stabilizing the sp\textsuperscript{3} bonds by thermally adjusting their local structures~\cite{Zeng_PhysRevB_109_214113_2024}.

\section{Conclusions}

In conclusion, our on-the-fly ML-augmented CAIMD simulations, which improve computational efficiency by approximately 99\% compared with conventional AIMD while maintaining an error of only \(\sim\)2.3\%, have elucidated several unresolved questions regarding the roles of pressure, temperature, uniaxial compression, and severe rotational shear strains in the structural transformation of GC.

We first simulate the tensile, compression, and shear deformation of GC using our developed CAIMD approach. The results reveal an unexpectedly high plasticity of GC in ambient conditions, particularly under large tensile and shear strains. Notably, strain-stiffening effects are observed in both compression and shear, contributing to the enhanced mechanical robustness of GC.

We then applied isotropic pressures up to 80~GPa to GC at various temperatures, followed by gradual decompression. As the temperature increases, annealing promotes substantial sp\textsuperscript{3} preservation and enables the formation of quenchable a-D. However, at 3000~K, graphitization becomes dominant, reducing the sp\textsuperscript{3} fraction to nearly zero and resulting in a density approaching that of graphite.

Further simulations mimicking experimental protocols that pressurizes GC at room temperature followed by heating, reveal a critical pressure threshold between 30--40~GPa, above which the sp\textsuperscript{3} percentage increases sharply with temperature. These simulations also demonstrate that the formation of a-D, which typically requires 60~GPa at room temperature, can instead be achieved by combining moderate pressure with elevated temperature. In particular, annealing the 40~GPa structure at 3000~K followed by cooling consistently yields quenchable a-D, where high pressures applied in advance can effectively avert high temperature graphitization.

In addition, GC transforms into a superhard phase under non-hydrostatic compression, capable of sustaining large stress differences between the uniaxial and confining directions. This stress difference increases markedly when the confining pressure exceeds 40~GPa, driven by a continuous rise in sp\textsuperscript{3} percentage induced by the applied uniaxial strain. In contrast, such behavior is not observed when the confining pressure is below 30~GPa.

Lastly, we investigated the influence of severe rotational shear strains on the phase transition behavior of GC. Our results indicate that shear strain can slightly lower the critical pressure threshold to approximately 30~GPa. At this pressure, shear strains exhibit an effect analogous to elevated temperatures, which significantly enhances the sp\textsuperscript{3} fraction, reaching up to 80\% even at room temperature. Unlike hydrostatic pressure, shear promotes more stable sp\textsuperscript{3} bonding, primarily because the resulting bond lengths are concentrated around those of cubic diamond under the same conditions. As a result, generally a sp\textsuperscript{3} content above 55\% is retained after unloading, in some cases approaching 80\%, with markedly reduced synthesizing temperature and pressure. 
These findings suggest that shear strain enhances the stability of sp\textsuperscript{3} bonding in GC and facilitates the formation of quenchable a-D under relatively low pressure and temperature.

Our findings validate the effectiveness of on-the-fly ML-augmented CAIMD simulations and clarify the distinct roles of pressure, temperature, uniaxial compression, and severe rotational shear strains in forming amorphous ultra-hard, sp\textsuperscript{3}-rich carbon. The strong agreement with experimental observations helps resolve key uncertainties in classical MD simulations, which are limited by fixed empirical potentials. Based on these advantages, our approach offers a robust framework for accurately modeling anisotropic stress effects in disordered and defective materials, with broad applicability to complex systems under extreme conditions.

\section*{Acknowledgments}  
This work was supported by the National Natural Science Foundation of China (Grant No. 11974237). Computations were performed at the Center for High Performance Computing, Shanghai Jiao Tong University.

\section*{Appendix A. Supplementary data}
Supplementary data to this article can be found online at arXiv.


\begin{thebibliography}{90}
\providecommand{\natexlab}[1]{#1}
\providecommand{\url}[1]{\texttt{#1}}
\expandafter\ifx\csname urlstyle\endcsname\relax
  \providecommand{\doi}[1]{doi: #1}\else
  \providecommand{\doi}{doi: \begingroup \urlstyle{rm}\Url}\fi

\bibitem[Behler and Parrinello(2007)]{Behler_PhysRevLett_98_146401_2007}
J.~Behler and M.~Parrinello.
\newblock Generalized neural-network representation of high-dimensional
  potential-energy surfaces.
\newblock \emph{Phys. Rev. Lett.}, 98:\penalty0 146401, 2007.
\newblock \doi{10.1103/PhysRevLett.98.146401}.

\bibitem[Bartók et~al.(2010)Bartók, Payne, Kondor, and
  et~al.]{Bartok_PhysRevLett_104_136403_2010}
A.~P. Bartók, M.~C. Payne, R.~Kondor, and et~al.
\newblock Gaussian approximation potentials: The accuracy of quantum mechanics,
  without the electrons.
\newblock \emph{Phys. Rev. Lett.}, 104:\penalty0 136403, 2010.
\newblock \doi{10.1103/PhysRevLett.104.136403}.

\bibitem[Smith et~al.(2019)Smith, Nebgen, Zubatyuk, and
  et~al.]{Smith_NatCommun_10_2903_2019}
J.~S. Smith, B.~T. Nebgen, R.~Zubatyuk, and et~al.
\newblock Approaching coupled cluster accuracy with a general-purpose neural
  network potential through transfer learning.
\newblock \emph{Nat. Commun.}, 10:\penalty0 2903, 2019.
\newblock \doi{10.1038/s41467-019-10827-4}.

\bibitem[Batzner et~al.(2022)Batzner, Musaelian, Sun, and
  et~al.]{Batzner_NatCommun_13_2453_2022}
S.~Batzner, A.~Musaelian, L.~Sun, and et~al.
\newblock E(3)-equivariant graph neural networks for data-efficient and
  accurate interatomic potentials.
\newblock \emph{Nat. Commun.}, 13:\penalty0 2453, 2022.
\newblock \doi{10.1038/s41467-022-29939-5}.

\bibitem[Daru et~al.(2022)Daru, Forbert, Behler, and
  et~al.]{Daru_PhysRevLett_129_226001_2022}
J.~Daru, H.~Forbert, J.~Behler, and et~al.
\newblock Coupled cluster molecular dynamics of condensed phase systems enabled
  by machine learning potentials: Liquid water benchmark.
\newblock \emph{Phys. Rev. Lett.}, 129:\penalty0 226001, 2022.
\newblock \doi{10.1103/PhysRevLett.129.226001}.

\bibitem[Pan et~al.(2024)Pan, Shi, Liang, and
  et~al.]{Pan_PhysRevB_110_224101_2024}
Shuning Pan, Jiuyang Shi, Zhixin Liang, and et~al.
\newblock Shock compression pathways to pyrite silica from machine learning
  simulations.
\newblock \emph{Phys. Rev. B}, 110:\penalty0 224101, 2024.
\newblock \doi{10.1103/PhysRevB.110.224101}.

\bibitem[Song et~al.(2024)Song, Zhao, Liu, and
  et~al.]{Song_NatCommun_15_10208_2024}
K.~Song, R.~Zhao, J.~Liu, and et~al.
\newblock General-purpose machine-learned potential for 16 elemental metals and
  their alloys.
\newblock \emph{Nat. Commun.}, 15:\penalty0 10208, 2024.
\newblock \doi{10.1038/s41467-024-54554-x}.

\bibitem[Wang et~al.(2024)Wang, Wang, Zhang, and
  et~al.]{Wang_NatCommun_15_7607_2024}
J.~Wang, Y.~Wang, H.~Zhang, and et~al.
\newblock E(n)-equivariant cartesian tensor message passing interatomic
  potential.
\newblock \emph{Nat. Commun.}, 15:\penalty0 7607, 2024.
\newblock \doi{10.1038/s41467-024-51886-6}.

\bibitem[Parrinello and Rahman(1980)]{Parrinello_PhysRevLett_45_1196_1980}
M.~Parrinello and A.~Rahman.
\newblock Crystal {Structure} and {Pair} {Potentials}: {A}
  {Molecular}-{Dynamics} {Study}.
\newblock \emph{Phys. Rev. Lett.}, 45:\penalty0 1196, 1980.
\newblock \doi{10.1103/PhysRevLett.45.1196}.

\bibitem[Kresse and Hafner(1993)]{Kresse_PhysRevB_47_558_1993}
G.~Kresse and J.~Hafner.
\newblock \textit{Ab initio} molecular dynamics for liquid metals.
\newblock \emph{Phys. Rev. B}, 47:\penalty0 558, 1993.
\newblock \doi{10.1103/PhysRevB.47.558}.

\bibitem[Kresse and Hafner(1994)]{Kresse_PhysRevB_49_14251_1994}
G.~Kresse and J.~Hafner.
\newblock \textit{{Ab} initio} molecular-dynamics simulation of the
  liquid-metal–amorphous-semiconductor transition in germanium.
\newblock \emph{Phys. Rev. B}, 49:\penalty0 14251, 1994.
\newblock \doi{10.1103/PhysRevB.49.14251}.

\bibitem[Kresse and Furthmüller(1996)]{Kresse_PhysRevB_54_11169_1996}
G.~Kresse and J.~Furthmüller.
\newblock Efficient iterative schemes for \textit{ab initio} total-energy
  calculations using a plane-wave basis set.
\newblock \emph{Phys. Rev. B}, 54:\penalty0 11169, 1996.
\newblock \doi{10.1103/PhysRevB.54.11169}.

\bibitem[Jinnouchi et~al.(2019)Jinnouchi, Karsai, and
  Kresse]{Jinnouchi_PhysRevB_100_014105_2019}
R.~Jinnouchi, F.~Karsai, and G.~Kresse.
\newblock On-the-fly machine learning force field generation: {Application} to
  melting points.
\newblock \emph{Phys. Rev. B}, 100:\penalty0 014105, 2019.
\newblock \doi{10.1103/PhysRevB.100.014105}.

\bibitem[Yamada and Sato(1962)]{Yamada_Nature_193_261_1962}
S.~Yamada and H.~Sato.
\newblock Some {Physical} {Properties} of {Glassy} {Carbon}.
\newblock \emph{Nature}, 193:\penalty0 261, 1962.
\newblock \doi{10.1038/193261b0}.

\bibitem[Yamaguchi(1963)]{Yamaguchi_Carbon_1_47_1963}
T.~Yamaguchi.
\newblock Galvanomagnetic properties of {Glassy} {Carbon}.
\newblock \emph{Carbon}, 1:\penalty0 47, 1963.
\newblock \doi{10.1016/0008-6223(63)90008-3}.

\bibitem[Cowlard and Lewis(1967)]{Cowlard_JMaterSci_2_507_1967}
F.~C. Cowlard and J.~C. Lewis.
\newblock Vitreous {Carbon} - {A} {New} {Form} of {Carbon}.
\newblock \emph{J. Mater. Sci.}, 2:\penalty0 507, 1967.
\newblock \doi{10.1007/BF00752216}.

\bibitem[Wang et~al.(2003)Wang, Zhang, Zhang, and
  et~al.]{Wang_Carbon_41_188_2003}
X.~Wang, G.~Zhang, Y.~Zhang, and et~al.
\newblock Graphitization of glassy carbon prepared under high temperatures and
  high pressures.
\newblock \emph{Carbon}, 41:\penalty0 188, 2003.
\newblock \doi{10.1016/S0008-6223(02)00319-6}.

\bibitem[Oishi et~al.(2014)Oishi, Botelho, Luscombe, and
  Rezende]{Oishi_Polimeros_24_541_2014}
S.~S. Oishi, E.~C. Botelho, C.~K. Luscombe, and M.~C. Rezende.
\newblock Synthesis and characterization of polyarylacetylene for use in the
  monolithic vitreous carbon processing.
\newblock \emph{Polimeros}, 24:\penalty0 541, 2014.
\newblock \doi{10.1590/0104-1428.1623}.

\bibitem[Botelho et~al.(2001)Botelho, Scherbakoff, and
  Rezende]{Botelho_Carbon_39_45_2001}
E.~C. Botelho, N.~Scherbakoff, and M.~C. Rezende.
\newblock Porosity control in glassy carbon by rheological study of the
  furfuryl resin.
\newblock \emph{Carbon}, 39:\penalty0 45, 2001.
\newblock \doi{10.1016/S0008-6223(00)00080-4}.

\bibitem[Oishi et~al.(2017)Oishi, Botelho, Rezende, and
  Ferreira]{Oishi_ApplSurfSci_394_87_2017}
S.~S. Oishi, E.~C. Botelho, M.~C. Rezende, and N.~G. Ferreira.
\newblock Structural and surface functionality changes in reticulated vitreous
  carbon produced from poly (furfuryl alcohol) with sodium hydroxide additions.
\newblock \emph{Appl. Surf. Sci.}, 394:\penalty0 87, 2017.
\newblock \doi{10.1016/j.apsusc.2016.10.112}.

\bibitem[Supplies()]{spi_supplies_2025}
SPI Supplies.
\newblock Glassy carbon.
\newblock \url{https://www.2spi.com/category/labware-crucibles-glassy-carbon/}.
\newblock Accessed 17 march 2025.

\bibitem[Co()]{als_co_2025}
A.L.S. Co.
\newblock Electrochemistry and spectroelectrochemistry, glassy carbon.
\newblock \url{https://www.als-japan.com/1408.html#defaultTab11}.
\newblock Accessed 17 march 2025.

\bibitem[GmbH()]{htw_glassy_2025}
H.T.W. Hochtemperatur-Werkstoffe GmbH.
\newblock Glassy carbon sigradur®.
\newblock \url{https://htw-germany.com/en/material}.
\newblock Accessed 17 march 2025.

\bibitem[Aldrich()]{merck_glassy_2025}
Merck-Sigma Aldrich.
\newblock Synlectrotm glassy carbon electrode.
\newblock \url{https://www.sigmaaldrich.cn/CN/zh/product/aldrich/esynth019}.
\newblock Accessed 17 March 2025.

\bibitem[GoodFellow()]{goodfellow_catalogue_2025}
GoodFellow.
\newblock Catalogue.
\newblock \url{https://www.goodfellow.com/}.
\newblock Accessed 17 March 2025.

\bibitem[Fielda and Swain(1996)]{Fielda_Carbon_34_1357_1996}
J.~S. Fielda and M.~V. Swain.
\newblock The indentation characterisation of the mechanical properties of
  various carbon materials: Glassy carbon, coke and pyrolytic graphite.
\newblock \emph{Carbon}, 34:\penalty0 1357, 1996.
\newblock \doi{10.1016/S0008-6223(96)00071-1}.

\bibitem[Gaefke et~al.(2007)Gaefke, Botelho, Ferreira, and
  Rezende]{Gaefke_JApplPolymSci_106_2274_2007}
C.~B. Gaefke, E.~C. Botelho, N.~G. Ferreira, and M.~C. Rezende.
\newblock Effect of furfuryl alcohol addition on the cure of furfuryl alcohol
  resin used in the glassy carbon manufacture.
\newblock \emph{J. Appl. Polym. Sci.}, 106:\penalty0 2274, 2007.
\newblock \doi{10.1002/app.26938}.

\bibitem[Uskoković(2021)]{Uskokovic_CarbonTrends_5_100116_2021}
V.~Uskoković.
\newblock A historical review of glassy carbon: {Synthesis}, structure,
  properties and applications.
\newblock \emph{Carbon Trends}, 5:\penalty0 100116, 2021.
\newblock \doi{10.1016/j.cartre.2021.100116}.

\bibitem[Vieira(2022)]{Vieira_Carbon_186_282_2022}
L.~de~S. Vieira.
\newblock A review on the use of glassy carbon in advanced technological
  applications.
\newblock \emph{Carbon}, 186:\penalty0 282, 2022.
\newblock \doi{10.1016/j.carbon.2021.10.022}.

\bibitem[Naseri et~al.(2023)Naseri, Niazi, Bagherzadeh, and
  et~al.]{Naseri_FoodChem_421_136195_2023}
M.~Naseri, A.~Niazi, K.~Bagherzadeh, and et~al.
\newblock Modified electrochemical aptasensor for ultrasensitive detection of
  tetracycline: In silico and in vitro studies.
\newblock \emph{Food Chem.}, 421:\penalty0 136195, 2023.
\newblock \doi{10.1016/j.foodchem.2023.136195}.

\bibitem[Cordeiro-Junior et~al.(2020)Cordeiro-Junior, Kronka, Goulart, and
  et~al.]{Cordeiro_JCatal_392_56_2020}
P.~J.~M. Cordeiro-Junior, M.~S. Kronka, L.~A. Goulart, and et~al.
\newblock Catalysis of oxygen reduction reaction for {H}$_2${O}$_2$
  electrogeneration: The impact of different conductive carbon matrices and
  their physicochemical properties.
\newblock \emph{J. Catal.}, 392:\penalty0 56, 2020.
\newblock \doi{10.1016/j.jcat.2020.09.020}.

\bibitem[Murray et~al.(2020)Murray, Recio, Caracciolo, and
  et~al.]{Murray_Carbon_167_388_2020}
V.~J. Murray, P.~Recio, A.~Caracciolo, and et~al.
\newblock Oxidation and nitridation of vitreous carbon at high temperatures.
\newblock \emph{Carbon}, 167:\penalty0 388, 2020.
\newblock \doi{10.1016/j.carbon.2020.05.076}.

\bibitem[Santos et~al.(2018)Santos, Montagna, Rezende, and
  Passador]{Santos_JApplPolymSci_136_47204_2018}
M.~S. Santos, L.~S. Montagna, M.~C. Rezende, and F.~R. Passador.
\newblock A new use for glassy carbon: development of {LDPE}/glassy carbon
  composites for antistatic packaging applications.
\newblock \emph{J. Appl. Polym. Sci.}, 136:\penalty0 47204, 2018.
\newblock \doi{10.1002/app.47204}.

\bibitem[Myalski et~al.(2020)Myalski, Godzierz, and
  Olesik]{Myalski_Polymers_12_2264_2020}
J.~Myalski, M.~Godzierz, and P.~Olesik.
\newblock Effect of carbon fillers on the wear resistance of {PA}6
  thermoplastic composites.
\newblock \emph{Polymers (Basel)}, 12:\penalty0 2264, 2020.
\newblock \doi{10.3390/polym12102264}.

\bibitem[Sure et~al.(2014)Sure, Shankar, Ramya, Mallika, and
  Mudali]{Sure_Carbon_67_643_2014}
J.~Sure, A.~R. Shankar, S.~Ramya, C.~Mallika, and U.~K. Mudali.
\newblock Corrosion behaviour of carbon materials exposed to molten lithium
  chloride-potassium chloride salt.
\newblock \emph{Carbon}, 67:\penalty0 643, 2014.
\newblock \doi{10.1016/j.carbon.2013.10.040}.

\bibitem[Robertson(2002)]{Robertson_MaterSciEngR_37_129_2002}
J.~Robertson.
\newblock Diamond-like amorphous carbon.
\newblock \emph{Mater. Sci. Eng., R}, 37:\penalty0 129, 2002.
\newblock \doi{10.1016/S0927-796X(02)00005-0}.

\bibitem[Erdemir and Donnet(2006)]{Erdemir_JPhysDApplPhys_39_R311_2006}
A.~Erdemir and C.~Donnet.
\newblock Tribology of diamond-like carbon films: Recent progress and future
  prospects.
\newblock \emph{J. Phys. D: Appl. Phys.}, 39:\penalty0 R311, 2006.
\newblock \doi{10.1088/0022-3727/39/18/R01}.

\bibitem[Hauert et~al.(2013)Hauert, Thorwarth, and
  Thorwarth]{Hauert_SurfCoatTechnol_233_119_2013}
R.~Hauert, K.~Thorwarth, and G.~Thorwarth.
\newblock An overview on diamond-like carbon coatings in medical applications.
\newblock \emph{Surf. Coat. Technol.}, 233:\penalty0 119, 2013.
\newblock \doi{10.1016/j.surfcoat.2013.04.015}.

\bibitem[Duan et~al.(2023)Duan, Yuan, Y., Yuan, and
  Tai]{Duan_JMaterChemC_11_5585_2023}
Z.~Duan, Z.~Yuan, Y., L.~Yuan, and H.~Tai.
\newblock Amorphous carbon material of daily carbon ink: emerging applications
  in pressure, strain, and humidity sensors.
\newblock \emph{J. Mater. Chem. C}, 11:\penalty0 5585, 2023.
\newblock \doi{10.1039/D3TC00016H}.

\bibitem[Bai et~al.(2016)Bai, Srikanth, Wu, and
  et~al.]{Bai_JNonCrystSolids_443_8_2016}
L.~Bai, N.~Srikanth, H.~Wu, and et~al.
\newblock Investigation on tensile behaviors of diamond-like carbon films.
\newblock \emph{J. Non-Cryst. Solids}, 443:\penalty0 8, 2016.
\newblock \doi{10.1016/j.jnoncrysol.2016.03.025}.

\bibitem[Xu et~al.(1997)Xu, Flynn, Tay, and et~al.]{Xu_PhilosMagB_76_351_1997}
S.~Xu, D.~Flynn, B.~K. Tay, and et~al.
\newblock Mechanical properties and raman spectra of tetrahedral amorphous
  carbon films with high sp\textsuperscript{3} fraction deposited using a
  filtered cathodic arc.
\newblock \emph{Philos. Mag. B}, 76:\penalty0 351, 1997.
\newblock \doi{10.1080/01418639708241099}.

\bibitem[McKenzie(1996)]{McKenzie_RepProgPhys_59_1611_1996}
D.~R. McKenzie.
\newblock Tetrahedral bonding in amorphous carbon.
\newblock \emph{Rep. Prog. Phys.}, 59:\penalty0 1611, 1996.
\newblock \doi{10.1088/0034-4885/59/12/002}.

\bibitem[Fallon et~al.(1993)Fallon, Veerasamy, Davis, and
  et~al.]{Fallon_PhysRevB_48_4777_1993}
P.~J. Fallon, V.~S. Veerasamy, C.~A. Davis, and et~al.
\newblock Properties of filtered-ion-beam-deposited diamondlike carbon as a
  function of ion energy.
\newblock \emph{Phys. Rev. B}, 48:\penalty0 4777, 1993.
\newblock \doi{10.1103/PhysRevB.48.4777}.

\bibitem[Pharr et~al.(1996)Pharr, Callahan, McAdams, and
  et~al.]{Pharr_ApplPhysLett_68_779_1996}
G.~M. Pharr, D.~L. Callahan, S.~D. McAdams, and et~al.
\newblock Hardness, elastic modulus, and structure of very hard carbon films
  produced by cathodic‐arc deposition with substrate pulse biasing.
\newblock \emph{Appl. Phys. Lett.}, 68:\penalty0 779, 1996.
\newblock \doi{10.1063/1.116530}.

\bibitem[Tan et~al.(2020)Tan, Sheng, Lou, and
  et~al.]{Tan_JPhysChemC_124_5489_2020}
L.-J. Tan, H.-W. Sheng, H.-B. Lou, and et~al.
\newblock High-pressure tetrahedral amorphous carbon synthesized by compressing
  glassy carbon at room temperature.
\newblock \emph{J. Phys. Chem. C}, 124:\penalty0 5489, 2020.
\newblock \doi{10.1021/acs.jpcc.0c00247}.

\bibitem[Klein et~al.(2016)Klein, Treske, Koitzsch, and
  et~al.]{Klein_Carbon_107_536_2016}
F.~Klein, U.~Treske, A.~Koitzsch, and et~al.
\newblock Nanoscale scanning electron microscopy based graphitization in
  tetrahedral amorphous carbon thin films.
\newblock \emph{Carbon}, 107:\penalty0 536, 2016.
\newblock \doi{10.1016/j.carbon.2016.06.002}.

\bibitem[Seok et~al.(2018)Seok, Kim, and Kim]{Seok_SciRep_8_13521_2018}
H.~J. Seok, J.~K. Kim, and H.~K. Kim.
\newblock Effective passivation of {Ag} nanowire network by transparent
  tetrahedral amorphous carbon film for flexible and transparent thin film
  heaters.
\newblock \emph{Sci. Rep.}, 8:\penalty0 13521, 2018.
\newblock \doi{10.1038/s41598-018-31927-z}.

\bibitem[Zhao et~al.(2024)Zhao, Lü, Chen, and
  et~al.]{Zhao_JApplPhys_135_065304_2024}
X.-W. Zhao, Y.-J. Lü, R.-B. Chen, and et~al.
\newblock The deposition properties of tetrahedral amorphous carbon coatings
  deposited on piston ring: Molecular dynamics simulation.
\newblock \emph{J. Appl. Phys.}, 135:\penalty0 065304, 2024.
\newblock \doi{10.1063/5.0189011}.

\bibitem[Lin et~al.(2011)Lin, Zhang, k.~Mao, and
  et~al.]{Lin_PhysRevLett_107_175504_2011}
Y.~Lin, L.~Zhang, H.~k.~Mao, and et~al.
\newblock Amorphous diamond: A high-pressure superhard carbon allotrope.
\newblock \emph{Phys. Rev. Lett.}, 107:\penalty0 175504, 2011.
\newblock \doi{10.1103/PhysRevLett.107.175504}.

\bibitem[Shiell et~al.(2016)Shiell, McCulloch, Bradby, and
  et~al.]{Shiell_SciRep_6_37232_2016}
T.~B. Shiell, D.~G. McCulloch, J.~E. Bradby, and et~al.
\newblock Nanocrystalline hexagonal diamond formed from glassy carbon.
\newblock \emph{Sci. Rep.}, 6:\penalty0 37232, 2016.
\newblock \doi{10.1038/srep37232}.

\bibitem[Zeng et~al.(2017)Zeng, Yang, Zeng, and
  et~al.]{Zeng_NatCommun_8_322_2017}
Z.-D. Zeng, L.-X. Yang, Q.-S. Zeng, and et~al.
\newblock Synthesis of quenchable amorphous diamond.
\newblock \emph{Nat. Commun.}, 8:\penalty0 322, 2017.
\newblock \doi{10.1038/s41467-017-00395-w}.

\bibitem[McCulloch et~al.(2020)McCulloch, Wong, Shiell, and
  et~al.]{McCulloch_Small_16_2004695_2020}
D.~G. McCulloch, S.~Wong, T.~B. Shiell, and et~al.
\newblock Investigation of room temperature formation of the ultra‐hard
  nanocarbons diamond and lonsdaleite.
\newblock \emph{Small}, 16:\penalty0 2004695, 2020.
\newblock \doi{10.1002/smll.202004695}.

\bibitem[Huang et~al.(2024)Huang, Shiell, Salek, and
  et~al.]{Huang_Carbon_219_118763_2024}
X.~Huang, T.~B. Shiell, A.~Salek, and et~al.
\newblock Comparison of hydrostatic and non-hydrostatic compression of glassy
  carbon to 80 {GP}a.
\newblock \emph{Carbon}, 219:\penalty0 118763, 2024.
\newblock \doi{10.1016/j.carbon.2023.118763}.

\bibitem[Solopova et~al.(2013)Solopova, Dubrovinskaia, and
  Dubrovinsky]{Solopova_ApplPhysLett_102_121909_2013}
N.~A. Solopova, N.~Dubrovinskaia, and L.~Dubrovinsky.
\newblock Raman spectroscopy of glassy carbon up to 60 {GPa}.
\newblock \emph{Appl. Phys. Lett.}, 102:\penalty0 121909, 2013.
\newblock \doi{10.1063/1.4798660}.

\bibitem[Yao et~al.(2014)Yao, Xiao, Fan, and
  et~al.]{Yao_ApplPhysLett_104_021916_2014}
M.-G. Yao, J.-P. Xiao, X.-H. Fan, and et~al.
\newblock Transparent, superhard amorphous carbon phase from compressing glassy
  carbon.
\newblock \emph{Appl. Phys. Lett.}, 104:\penalty0 021916, 2014.
\newblock \doi{10.1063/1.4861929}.

\bibitem[Yao et~al.(2017)Yao, Fan, Zhang, and
  et~al.]{Yao_ApplPhysLett_111_101901_2017}
M.-G. Yao, X.-H. Fan, W.-W. Zhang, and et~al.
\newblock Uniaxial-stress-driven transformation in cold compressed glassy
  carbon.
\newblock \emph{Appl. Phys. Lett.}, 111:\penalty0 101901, 2017.
\newblock \doi{10.1063/1.4996278}.

\bibitem[Shiell et~al.(2018)Shiell, McCulloch, McKenzie, and
  et~al.]{Shiell_PhysRevLett_120_215701_2018}
T.~B. Shiell, D.~G. McCulloch, D.~R. McKenzie, and et~al.
\newblock Graphitization of {Glassy} {Carbon} after {Compression} at {Room}
  {Temperature}.
\newblock \emph{Phys. Rev. Lett.}, 120:\penalty0 215701, 2018.
\newblock \doi{10.1103/PhysRevLett.120.215701}.

\bibitem[Zeng et~al.(2024)Zeng, Sheng, Lou, and
  et~al.]{Zeng_PhysRevB_109_214113_2024}
Z.-D. Zeng, H.-W. Sheng, H.-B. Lou, and et~al.
\newblock Mechanism of thermally assisted stabilization of pressure-induced
  $s{p}^{3}$ bonds in amorphous carbon.
\newblock \emph{Phys. Rev. B}, 109:\penalty0 214113, 2024.
\newblock \doi{10.1103/PhysRevB.109.214113}.

\bibitem[Levitas(2023)]{Levitas_MaterialsTrans_64_1866_2023}
Valery~I. Levitas.
\newblock Recent \textit{In Situ} experimental and theoretical advances in
  severe plastic deformations, strain-induced phase transformations, and
  microstructure evolution under high pressure.
\newblock \emph{Mater. Trans.}, 64:\penalty0 1866, 2023.
\newblock \doi{10.2320/matertrans.MT-MF2022055}.

\bibitem[Edalati(2024)]{Edalati_MaterialsTrans_65_466_2024}
Kaveh Edalati.
\newblock Severe plastic deformation of light metals (magnesium, aluminum and
  titanium) and alloys by high-pressure torsion: Review of fundamentals and
  mechanical/functional properties.
\newblock \emph{Mater. Trans.}, 65:\penalty0 466, 2024.
\newblock \doi{10.2320/matertrans.MT-L2023022}.

\bibitem[Gao et~al.(2019)Gao, Ma, An, and et. al.]{Gao_Carbon_146_364_2019}
Y.~Gao, Y.-Z. Ma, Q.~An, and et. al.
\newblock Shear driven formation of nano-diamonds at sub-gigapascals and
  300~{K}.
\newblock \emph{Carbon}, 146:\penalty0 364, 2019.
\newblock \doi{10.1016/j.carbon.2019.02.012}.

\bibitem[Yang et~al.(2025)Yang, Yuan, Qian, and
  et~al.]{Yang_Carbon_232_119802_2025}
Y.~Yang, M.-Z. Yuan, C.~Qian, and et~al.
\newblock Disordering of graphene nanoplatelet, carbon nanotube and {C}$_{60}$
  fullerene under shear stress.
\newblock \emph{Carbon}, 232:\penalty0 119802, 2025.
\newblock \doi{10.1016/j.carbon.2024.119802}.

\bibitem[Zeng et~al.(2019)Zeng, Sheng, Yang, and
  et~al.]{Zeng_PhysRevMater_3_033608_2019}
Z.-D. Zeng, H.-W. Sheng, L.-X. Yang, and et~al.
\newblock Structural transition in cold-compressed glassy carbon.
\newblock \emph{Phys. Rev. Mater.}, 3:\penalty0 033608, 2019.
\newblock \doi{10.1103/PhysRevMaterials.3.033608}.

\bibitem[Jenkins and Kawamura(1971)]{Jenkins_Nature_231_175_1971}
G.~M. Jenkins and K.~Kawamura.
\newblock Structure of glassy carbon.
\newblock \emph{Nature}, 231:\penalty0 175, 1971.
\newblock \doi{10.1038/231175a0}.

\bibitem[Yoshida et~al.(1991)Yoshida, Kaburagi, and
  Hishiyama]{Yoshida_Carbon_29_1107_1991}
A.~Yoshida, Y.~Kaburagi, and Y.~Hishiyama.
\newblock Microtexture and magnetoresistance of glass-like carbons.
\newblock \emph{Carbon}, 29:\penalty0 1107, 1991.
\newblock \doi{10.1016/0008-6223(91)90027-G}.

\bibitem[Harris(2004)]{Harris_PhilosMagA_84_3159_2004}
P.~J.~F. Harris.
\newblock Fullerene-related structure of commercial glassy carbons.
\newblock \emph{Philos. Mag. A}, 84:\penalty0 3159, 2004.
\newblock \doi{10.1080/14786430410001720363}.

\bibitem[Jurkiewicz et~al.(2017)Jurkiewicz, Duber, Fischerd, and
  Burian]{Jurkiewicz_JApplCrystallogr_50_36_2017}
K.~Jurkiewicz, S.~Duber, H.~E. Fischerd, and A.~Burian.
\newblock Modelling of glass-like carbon structure and its experimental
  verification by neutron and x-ray diffraction.
\newblock \emph{J. Appl. Crystallogr.}, 50:\penalty0 36, 2017.
\newblock \doi{10.1107/S1600576716017660}.

\bibitem[Montgomery-Walsh et~al.(2021)Montgomery-Walsh, Nimbalkar, Bunnell, and
  et~al.]{MontgomeryWalsh_Carbon_184_627_2021}
R.~Montgomery-Walsh, S.~Nimbalkar, J.~Bunnell, and et~al.
\newblock Molecular dynamics simulation of evolution of nanostructures and
  functional groups in glassy carbon under pyrolysis.
\newblock \emph{Carbon}, 184:\penalty0 627, 2021.
\newblock \doi{10.1016/j.carbon.2021.08.070}.

\bibitem[Wen and Sun(2018)]{Wen_PhysRevB_98_014103_2018}
L.-B. Wen and H.~Sun.
\newblock Understanding shear-induced
  sp\textsuperscript{2}-to-sp\textsuperscript{3} phase transitions in glassy
  carbon at low pressure using first-principles calculations.
\newblock \emph{Phys. Rev. B}, 98:\penalty0 014103, 2018.
\newblock \doi{10.1103/PhysRevB.98.014103}.

\bibitem[Jiang et~al.(2013)Jiang, Århammar, Liu, and
  et~al.]{Jiang_SciRep_3_1877_2013}
X.~Jiang, C.~Århammar, P.~Liu, and et~al.
\newblock The {R}3-carbon allotrope: a pathway towards glassy carbon under high
  pressure.
\newblock \emph{Sci. Rep.}, 3:\penalty0 1877, 2013.
\newblock \doi{10.1038/srep01877}.

\bibitem[de~Tomas et~al.(2016)de~Tomas, Suarez-Martinez, and
  Marks]{Tomas_Carbon_109_681_2016}
C.~de~Tomas, I.~Suarez-Martinez, and N.~A. Marks.
\newblock Graphitization of amorphous carbons: A comparative study of
  interatomic potentials.
\newblock \emph{Carbon}, 109:\penalty0 681, 2016.
\newblock \doi{10.1016/j.carbon.2016.08.024}.

\bibitem[Hossain et~al.(2021)Hossain, Zhang, Cheng, and
  et~al.]{Hossain_Carbon_183_940_2021}
Md~D. Hossain, Q.~Zhang, T.~Cheng, and et~al.
\newblock Graphitization of low-density amorphous carbon for electrocatalysis
  electrodes from {ReaxFF} reactive dynamics.
\newblock \emph{Carbon}, 183:\penalty0 940, 2021.
\newblock \doi{10.1016/j.carbon.2021.07.080}.

\bibitem[Yeh et~al.(2025)Yeh, Tran, and
  D.~B.~Knorr]{Yeh_Carbon_234_120006_2025}
I.-C. Yeh, N.~T. Tran, and Jr. D.~B.~Knorr.
\newblock Effects of high-temperature annealing on structural and mechanical
  properties of amorphous carbon materials investigated by molecular dynamics
  simulations.
\newblock \emph{Carbon}, 234:\penalty0 120006, 2025.
\newblock \doi{10.1016/j.carbon.2025.120006}.

\bibitem[Hu et~al.(2021)Hu, Zhang, Liu, and et~al.]{Hu_JMateriomics_7_177_2021}
M.~Hu, S.-S. Zhang, B.~Liu, and et~al.
\newblock Heat-treated glassy carbon under pressure exhibiting superior
  hardness, strength and elasticity.
\newblock \emph{J. Materiomics}, 7:\penalty0 177, 2021.
\newblock \doi{10.1016/j.jmat.2020.06.007}.

\bibitem[Kresse and Joubert(1999)]{Kresse_PhysRevB_59_1758_1999}
G.~Kresse and D.~Joubert.
\newblock From ultrasoft pseudopotentials to the projector augmented-wave
  method.
\newblock \emph{Phys. Rev. B}, 59:\penalty0 1758, 1999.
\newblock \doi{10.1103/PhysRevB.59.1758}.

\bibitem[Perdew et~al.(2008)Perdew, Ruzsinszky, Csonka, and
  et~al.]{Perdew_PhysRevLett_100_136406_2008}
J.~P. Perdew, A.~Ruzsinszky, G.~I. Csonka, and et~al.
\newblock Restoring the density-gradient expansion for exchange in solids and
  surfaces.
\newblock \emph{Phys. Rev. Lett.}, 100:\penalty0 136406, 2008.
\newblock \doi{10.1103/PhysRevLett.100.136406}.

\bibitem[Perdew et~al.(1996)Perdew, Burke, and
  Ernzerhof]{Perdew_PhysRevLett_77_3865_1996}
J.~P. Perdew, K.~Burke, and M.~Ernzerhof.
\newblock Generalized gradient approximation made simple.
\newblock \emph{Phys. Rev. Lett.}, 77:\penalty0 3865, 1996.
\newblock \doi{10.1103/PhysRevLett.77.3865}.

\bibitem[Monkhorst and Pack(1976)]{Monkhorst_PhysRevB_13_5188_1976}
H.~J. Monkhorst and J.~D. Pack.
\newblock Special points for brillouin-zone integrations.
\newblock \emph{Phys. Rev. B}, 13:\penalty0 5188, 1976.
\newblock \doi{10.1103/PhysRevB.13.5188}.

\bibitem[Marks(1997)]{Marks_PhysRevB_56_2441_1997}
N.~A. Marks.
\newblock Evidence for subpicosecond thermal spikes in the formation of
  tetrahedral amorphous carbon.
\newblock \emph{Phys. Rev. B}, 56:\penalty0 2441, 1997.
\newblock \doi{10.1103/PhysRevB.56.2441}.

\bibitem[Parrinello and Rahman(1981)]{Parrinello_JApplPhys_52_7182_1981}
M.~Parrinello and A.~Rahman.
\newblock Polymorphic transitions in single crystals: A new molecular dynamics
  method.
\newblock \emph{J. Appl. Phys.}, 52:\penalty0 7182, 1981.
\newblock \doi{10.1063/1.328693}.

\bibitem[Wen et~al.(2019)Wen, Wu, Sun, and Chen]{Wen_Carbon_155_361_2019}
L.-B. Wen, H.~Wu, H.~Sun, and C.-F. Chen.
\newblock Profound softening and shear-induced melting of diamond under extreme
  conditions: An \textit{ab-initio} molecular dynamics study.
\newblock \emph{Carbon}, 155:\penalty0 361, 2019.
\newblock \doi{10.1016/j.carbon.2019.08.079}.

\bibitem[Yang and Sun(2023)]{Yang_PhysRevB_107_104101_2023}
M.~Yang and H.~Sun.
\newblock Early melting of tantalum carbide under anisotropic stresses: An
  \textit{ab initio} molecular dynamics study.
\newblock \emph{Phys. Rev. B}, 107:\penalty0 104101, 2023.
\newblock \doi{10.1103/PhysRevB.107.104101}.

\bibitem[Cheng and Sun(2024)]{Cheng_PhysRevMater_8_113604_2024}
M.-Q. Cheng and H.~Sun.
\newblock Stability of c-{BAs} under extreme conditions: {An} \textit{ab
  initio} molecular dynamics study.
\newblock \emph{Phys. Rev. Mater.}, 8:\penalty0 113604, 2024.
\newblock \doi{10.1103/PhysRevMaterials.8.113604}.

\bibitem[Ma et~al.(2006)Ma, Levitas, and
  Hashemi]{Ma_JPhysChemSolids_67_2083_2006}
Y.-Z. Ma, V.~I. Levitas, and J.~Hashemi.
\newblock X-ray diffraction measurements in a rotational diamond anvil cell.
\newblock \emph{J. Phys. Chem. Solids}, 67:\penalty0 2083, 2006.
\newblock \doi{10.1016/j.jpcs.2006.05.052}.

\bibitem[Ciezak and Jenkins(2011)]{Ciezak_RevSciInstrum_82_073905_2011}
J.~A. Ciezak and T.~A. Jenkins.
\newblock Optical cell for \textit{in situ} vibrational spectroscopic
  measurements at high pressures and shear.
\newblock \emph{Rev. Sci. Instrum.}, 82:\penalty0 073905, 2011.
\newblock \doi{10.1063/1.3606640}.

\bibitem[de~Tomas et~al.(2018)de~Tomas, Suarez-Martinez, and
  Marks]{Tomas_ApplPhysLett_112_251907_2018}
C.~de~Tomas, I.~Suarez-Martinez, and N.~A. Marks.
\newblock Carbide-derived carbons for dense and tunable 3{D} graphene networks.
\newblock \emph{Appl. Phys. Lett.}, 112:\penalty0 251907, 2018.
\newblock \doi{10.1063/1.5030136}.

\bibitem[Banerjee et~al.(2018)]{Banerjee_Science_360_300_2018}
A.~Banerjee et~al.
\newblock Ultralarge elastic deformation of nanoscale diamond.
\newblock \emph{Science}, 360:\penalty0 300, 2018.
\newblock \doi{10.1126/science.aar4165}.

\bibitem[Dang et~al.(2021)]{Dang_Science_371_76_2021}
C.~Dang et~al.
\newblock Achieving large uniform tensile elasticity in microfabricated
  diamond.
\newblock \emph{Science}, 371:\penalty0 76, 2021.
\newblock \doi{10.1126/science.abc4174}.

\bibitem[Shiell et~al.(2019)Shiell, de~Tomas, McCulloch, and
  et~al.]{Shiell_PhysRevB_99_024114_2019}
T.~B. Shiell, C.~de~Tomas, D.~G. McCulloch, and et~al.
\newblock \textit{In situ} analysis of the structural transformation of glassy
  carbon under compression at room temperature.
\newblock \emph{Phys. Rev. B}, 99:\penalty0 024114, 2019.
\newblock \doi{10.1103/PhysRevB.99.024114}.

\bibitem[Wong et~al.(2019)Wong, Shiell, Cook, and
  et~al.]{Wong_Carbon_142_475_2019}
S.~Wong, T.~B. Shiell, B.~A. Cook, and et~al.
\newblock The shear-driven transformation mechanism from glassy carbon to
  hexagonal diamond.
\newblock \emph{Carbon}, 142:\penalty0 475, 2019.
\newblock \doi{10.1016/j.carbon.2018.10.080}.

\end{thebibliography}

\FloatBarrier

\end{document}


\title{{\Large Supplemental Materials for}\\\vbox{} 
		``On-the-fly machine learning-augmented constrained AIMD 
		to design new routes from glassy carbon to quenchable amorphous diamond with low pressure and temperature"} 
	
	\author{Meng-Qi Cheng}
	\author{Wei-Dong Luo}
	
	\author{Hong Sun}
	\email{hsun@sjtu.edu.cn}

	\affiliation{%
		School of Physics and Astronomy and Key Laboratory of Artificial Structures and Quantum Control (Ministry of Education), Shanghai Jiao Tong University, Shanghai 200240, China\\
		}%

	\date{\today}

	\maketitle 
	
\section{Fundamental Principles of Constrained AIMD}

\setlength{\parskip}{4pt}

To implement constrained \textit{ab initio} molecular dynamics (AIMD) simulations within the \textit{NpT} ensemble, we modified the source code of the Vienna Ab initio Simulation Package (VASP). Given that VASP's \textit{NpT} AIMD calculations are fundamentally based on the methodology originally proposed by Parrinello and Rahman, we outline our modifications using their established theoretical framework. Many equations presented herein are directly derived from their seminal works—\textit{Phys. Rev. Lett.} \textbf{45}, 1196 (1980) and \textit{J. Appl. Phys.} \textbf{52}, 7182 (1981)—with particular emphasis on the formulations presented in the latter reference. Definitions and notations of various physical quantities follow strictly to those in these references, and are not restated here except where necessary for clarity.

We start by considering a system of \( N \) atoms contained within a supercell defined by lattice vectors \((\mathbf{a}, \mathbf{b}, \mathbf{c})\), collectively represented as the \(3 \times 3\) matrix \( \mathbf{h} \). When subjected to an external hydrostatic pressure \( p \) and an anisotropic stress tensor \( \mathbf{S} \), the dynamics of this system—encompassing \( 3N + 9 \) degrees of freedom—is described by the following Lagrangian:

\begin{equation}\label{1}
	\mathscr{L} = \frac{1}{2} \sum_{i=1}^{N} m_{i} \dot{\mathbf{s}}_{i}^{\prime} \mathbf{G} \dot{\mathbf{s}}_{i}
	- \sum_{i=1}^{N} \sum_{j>i}^{N} \phi(r_{ij})
	+ \frac{1}{2} W \operatorname{Tr}(\dot{\mathbf{h}}^{\prime} \dot{\mathbf{h}})
	- p \Omega
	- \frac{1}{2} \operatorname{Tr}(\boldsymbol{\Sigma} \mathbf{G})
\end{equation}

Here, \( m_i \) denotes the atomic mass, \( \mathbf{s}_i \) are fractional atomic coordinates, \( \mathbf{G} = \mathbf{h}'\mathbf{h} \) is the metric tensor, \( \phi(r_{ij}) \) is the interatomic potential, \( W \) is the fictitious mass associated with lattice degrees of freedom, \( \Omega \) represents the supercell volume, and \( \boldsymbol{\Sigma} \) is directly related to the externally applied stress tensor \( \mathbf{S} \).

Applying the Euler-Lagrange equations to this Lagrangian yields the equations of motion governing atomic positions and lattice dynamics:

\begin{equation}\label{2}
	\ddot{\mathbf{s}}_{i} = -\sum_{j \neq i} m_{i}^{-1} \frac{\phi^{\prime}(r_{ij})}{r_{ij}} (\mathbf{s}_{i}-\mathbf{s}_{j}) - \mathbf{G}^{-1} \dot{\mathbf{G}} \dot{\mathbf{s}}_i \quad (i=1,2, \ldots, N)
\end{equation}

\begin{equation}\label{3}
	W \ddot{\mathbf{h}} = (\boldsymbol{\pi}-p)\boldsymbol{\sigma} - \mathbf{h} \boldsymbol{\Sigma}
\end{equation}

These equations are numerically integrated using standard MD algorithms such as the velocity Verlet method, with careful timestep control to ensure numerical stability and accurate simulation results.
In standard VASP implementations of the \textit{NpT} ensemble, the anisotropic stress term represented by \(\frac{1}{2}\operatorname{Tr}(\boldsymbol{\Sigma}\mathbf{G})\) in the Lagrangian is typically not considered explicitly. In our modification, this term plays a crucial role, enabling the direct application and control of anisotropic stress or strain during constrained AIMD simulations.

To illustrate the implementation of constrained molecular dynamics within the \textit{NpT} ensemble, we consider a uniaxial compression along the $z$ axis direction as an example. The deformation of the lattice \( \mathbf{h} \), defined by lattice vectors \(\mathbf{a}, \mathbf{b}, \mathbf{c}\), is constrained between time steps \(n\) and \(n+1\) during the AIMD simulations, with a timestep \(\Delta t = 1~\text{fs}\)
and strain matrix $\Delta \boldsymbol{\epsilon}$, as follows:

\begin{equation}\label{4} 
	\mathbf{h}^{(n+1)}=\mathbf{h}^{(n)} \times \Delta \boldsymbol{\epsilon} ,
	\quad \text{where} \quad
	\Delta \boldsymbol{\epsilon}=
	\begin{bmatrix}
		1 & 0 & 0 \\
		0 & 1 & 0 \\
		0 & 0 & 1-\dot{\varepsilon}\Delta t
	\end{bmatrix};   \quad
	\mathbf{h}^{(n)}=
	\begin{bmatrix}
		a_{1}^{(n)} & a_{2}^{(n)} & a_{3}^{(n)} \\
		b_{1}^{(n)} & b_{2}^{(n)} & b_{3}^{(n)} \\
		c_{1}^{(n)} & c_{2}^{(n)} & c_{3}^{(n)}
	\end{bmatrix}.
\end{equation}
Expanding explicitly, the deformation constraints yield:

\begin{equation}\label{5}
	a_{3}^{(n+1)} - a_{3}^{(n)} = -\dot{\varepsilon} \Delta t\,a_{3}^{(n)}, \quad 
	b_{3}^{(n+1)} - b_{3}^{(n)} = -\dot{\varepsilon} \Delta t\,b_{3}^{(n)}, \quad 
	c_{3}^{(n+1)} - c_{3}^{(n)} = -\dot{\varepsilon} \Delta t\,c_{3}^{(n)}.
\end{equation}
These discrete relations can be represented by their continuous analogs:

\begin{equation}\label{6}
	\frac{da_{3}}{dt} = -\dot{\varepsilon}\,a_{3}, \quad 
	\frac{db_{3}}{dt} = -\dot{\varepsilon}\,b_{3}, \quad 
	\frac{dc_{3}}{dt} = -\dot{\varepsilon}\,c_{3},
\end{equation}
where \(\dot{\varepsilon}\) is the imposed strain rate (set to \(0.001~\text{ps}^{-1}\) in our simulations).

Under these constraints, the dynamical equations for the lattice vectors \(a_3, b_3, c_3\), originally described by Eq.~\eqref{3}, are replaced by the constraints defined in Eq.~\eqref{6}. Specifically, this replacement sets the relaxation of these lattice vector components to zero, thereby directly controlling the uniaxial compressive strain. Consequently, the resulting stress component \(S_{zz}\) emerges naturally from this constrained deformation. All other stress components in the tensor \(\mathbf{S}\) remain free to relax to zero, governed by the unconstrained components of Eq.~\eqref{3}.

This constrained simulation approach allows precise control of deformation while accurately capturing stress-strain relationships within a computationally efficient AIMD framework.

\newpage
\section{Simulation of rotational shear in RDAC by CAIMD}
\begin{figure}[H]
	\centering
	\includegraphics[width=0.7\textwidth]{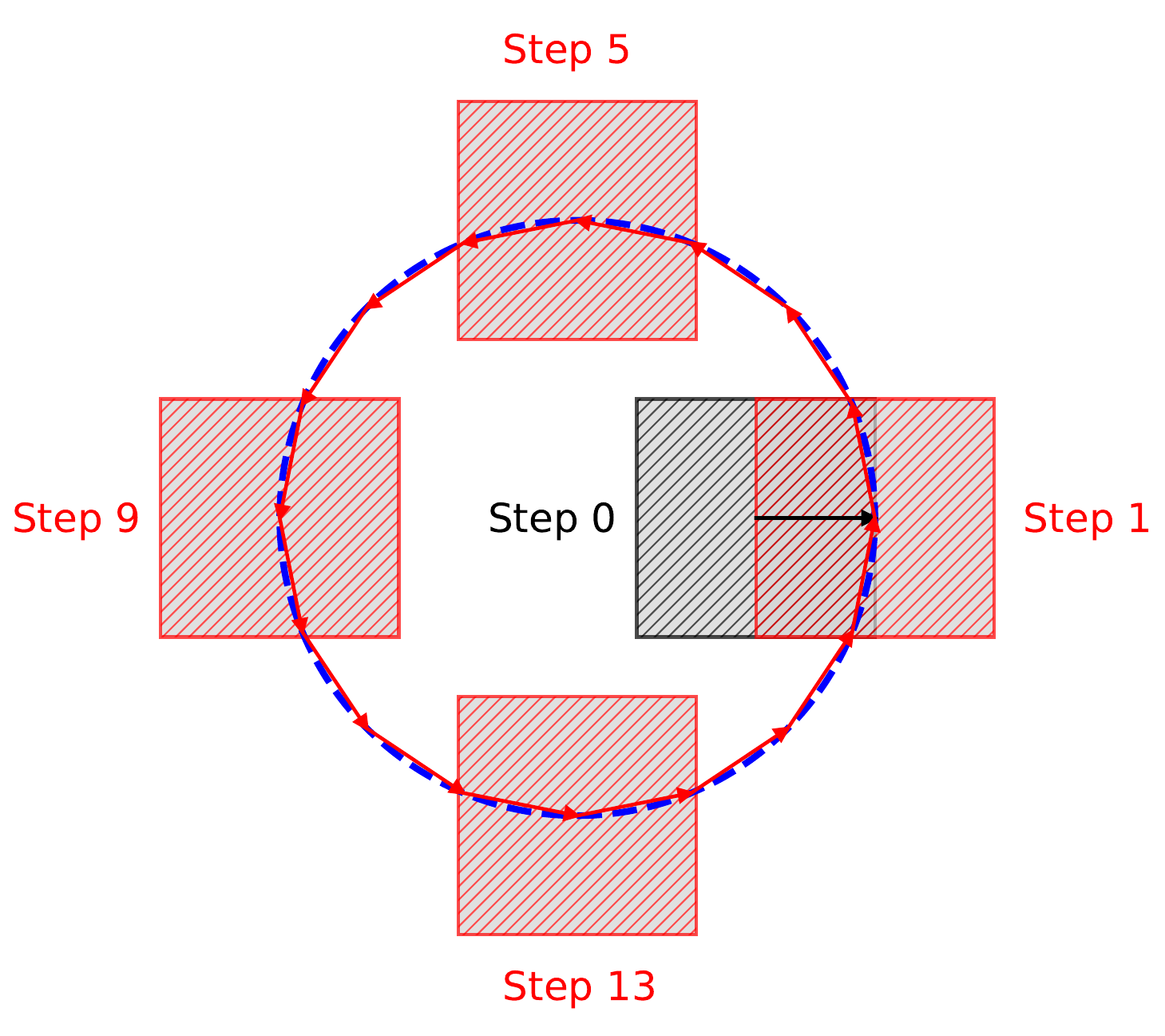}
	\caption{The schematic illustration of rotational shear. The result is shown in the \(x\)--\(y\) plane, representing a top view of the GC supercell structure.
	 }
	\label{RADAC}
\end{figure}

Fig.~\ref{RADAC} presents a schematic illustration of the rotational shear process conducted within a plane normal to the \(z\) axis. 
The procedure begins by applying a linear shear strain of 0.2 in an arbitrary in-plane direction, which displaces the GC supercell from its equilibrium configuration by shear moving the top layer of the supercell (the red square) away from its bottom layer (the black square) along the black arrow, labeled as Step~0. This is followed by 16 consecutive rotational shear steps, represented by the red square labeled as Step~1-16, with the shear directions indicated by the red arrows rotated by 22.5\(^\circ\) in each step, thereby completing a full 360\(^\circ\) rotation, while the bottom layer (the black square) of the supercell remains fixed. 
Each step is executed at a constant shear strain rate of 0.001~ps\(^{-1}\) with a time step of 1~fs, requiring 200{,}000 constrained AIMD (CAIMD) steps per direction. 
In total, 3{,}400{,}000 CAIMD steps (equivalent to 3400~ps) are used to complete the full rotational shear cycle.

\newpage
\section{Comparison of on-the-fly ML-augmented CAIMD and Exact CAIMD}
A comprehensive comparison between on-the-fly machine learning (ML)-augmented constrained AIMD (CAIMD) and conventional CAIMD is illustrated through simulations of cubic boron arsenide (c-BAs) undergoing compression and shear deformation at 300~K. Our computational simulations were conducted using the AIMD capabilities of the VASP, employing the projector augmented-wave pseudopotential and the Perdew-Burke-Ernzerhof generalized gradient approximation exchange-correlation functional. A plane-wave basis energy cutoff of 520~\text{eV} and a \( 2 \times 2 \times 2\) Monkhorst-Pack K-point grid were adopted for a 64-atom c-BAs supercell, under periodic boundary conditions. To implement CAIMD simulations for compression (or tensile) deformations, we align the direction \(\left[klm\right]\) of applied strains with the \(x\)-axis, and for linear shear deformations, we set the shearing plane \(\left(pqr\right)\) normal to the \(z\)-axis, while the shearing direction \(\left[klm\right]\) aligns parallel to the \(x\)-axis.

\begin{figure}[H]
	\centering
	\includegraphics[width=0.8\textwidth]{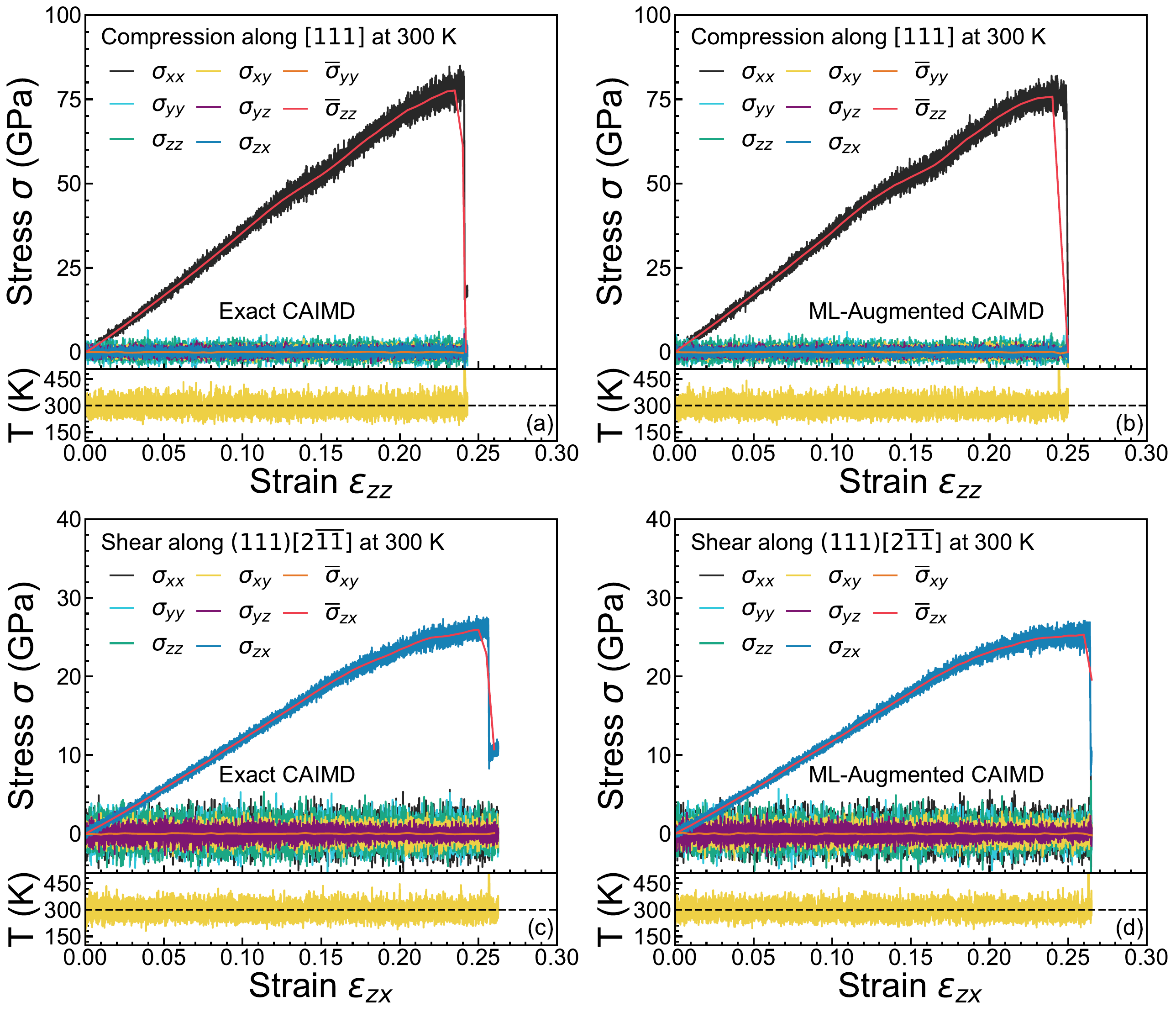}
	\caption{Comparison between conventional (a,c) and ML-augmented (b,d) CAIMD simulations of c-BAs at 300~K. Upper panels: Instantaneous fluctuations of the six stress components as strains increase linearly with CAIMD timesteps, alongside representative average stress curves (e.g., \(\overline{\sigma}_{xx}\), \(\overline{\sigma}_{zx}\)). Lower panels: Temperature evolution during the compression and shear processes at 300 K.}
	\label{c-BAs}
\end{figure}

As demonstrated in Fig.\ref{c-BAs}, we compare the outcomes for compression along \(\left[111\right]\) and shear along \(\left(111\right)\left[2\overline{11}\right]\) directions. The stress-strain relationship from on-the-fly ML-augmented CAIMD simulations closely matches that from conventional CAIMD. Specifically, for compression along the \(\left[111\right]\) direction, the maximum stress decreases by 2.3\%, from 77.6~GPa to 75.8~GPa, while the corresponding strain increases from 0.235 to 0.24. For shear along the \(\left(111\right)\left[2\overline{11}\right]\) direction, the maximum stress also decreases by 2.3\%, from 25.9~GPa to 25.3~GPa, with the corresponding strain increasing from 0.25 to 0.26.
Importantly, maintaining such high accuracy, approximately 99\% of the ionic steps in the ML-augmented CAIMD utilized on-the-fly machine-learned potentials under the same computational resources (two computing nodes totaling 128 cores). This resulted in a computational efficiency improvement of \(\sim\)99\%.

\newpage
\section{Tensile Structural Evolution Snapshots under 0~GPa at 300~K}
\begin{figure}[H]
	\centering
	\includegraphics[width=1\textwidth]{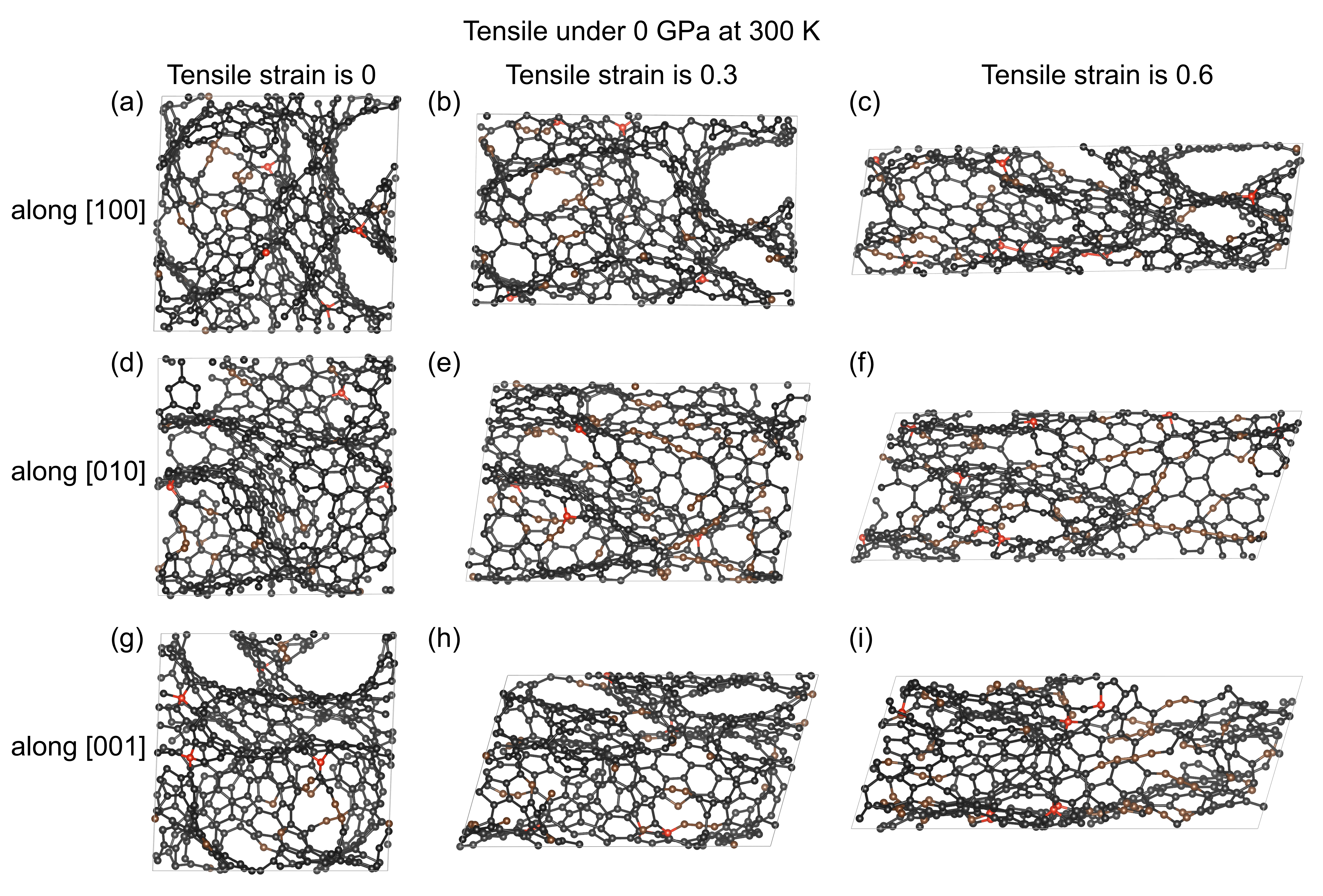}
	\caption{Atomic structures of GC at 0~GPa and 300~K under uniaxial tensile strains of 0, 0.3, and 0.6 applied along the [100] (a–c), [010] (d–f), and [001] (g–i) directions. Brown-colored atoms represent sp-bonded carbon, gray-colored atoms indicate sp\textsuperscript{2}-bonded carbon, and red-colored atoms correspond to sp\textsuperscript{3}-bonded carbon.}
	\label{Tensile_snapshots}
\end{figure}

\newpage
\section{Pressure Dependence of Density and sp\textsuperscript{3}-Bonds at 2900~K}
\begin{figure}[H]
	\centering
	\includegraphics[width=0.43\textwidth]{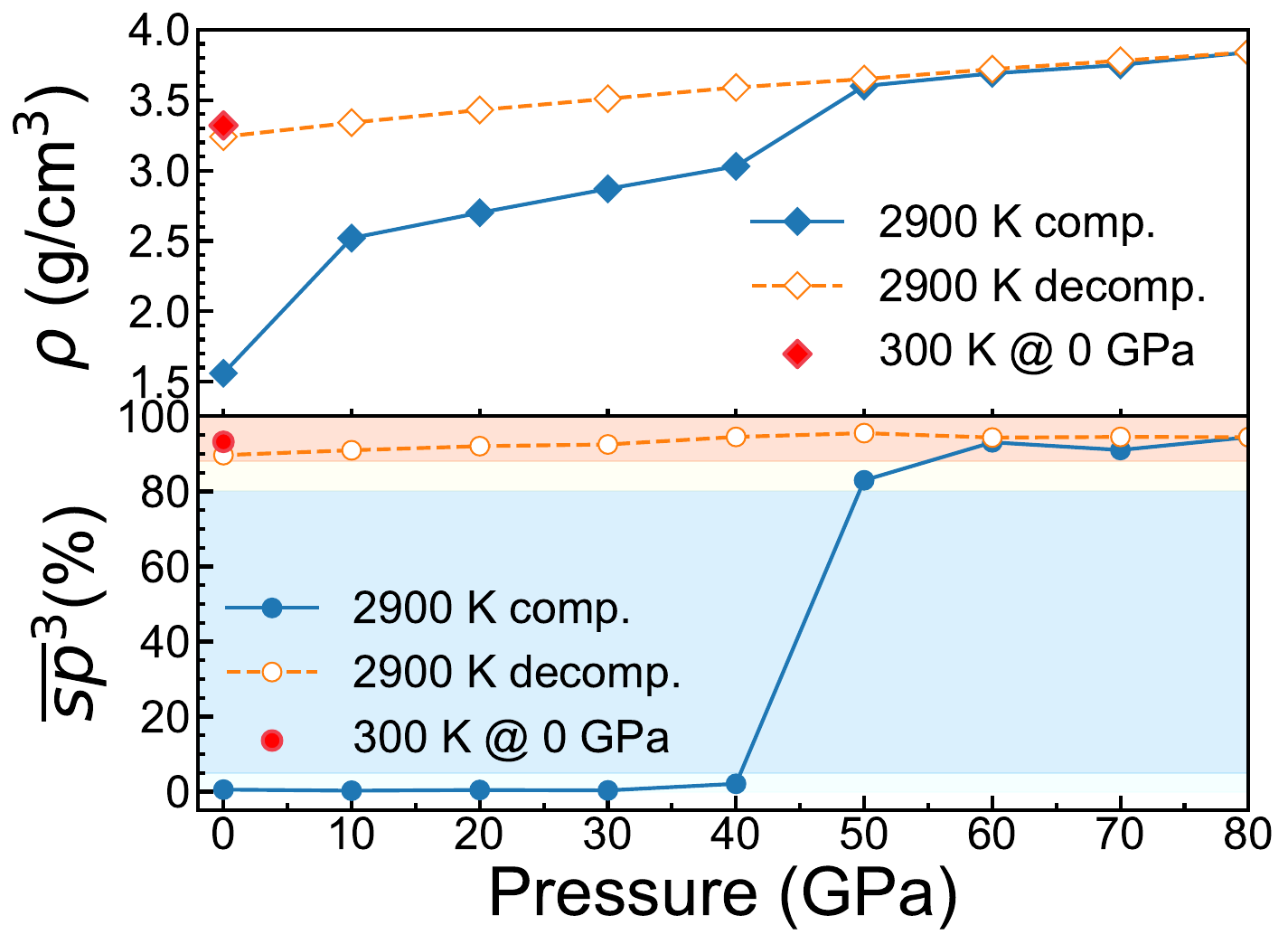}
	\caption{The average density (upper panel) and the percentage of sp\textsuperscript{3} bonds (lower panel) as a function of pressure at 2900~K. Each data point represents the time-averaged value obtained from at least 50~ps of simulation after equilibration. Solid lines indicate stepwise compression, while dashed lines represent stepwise decompression. The red sample marks the final density or sp\textsuperscript{3}-bond percentage at 300~K and 0~GPa. The percentage of sp\textsuperscript{3} bonds is categorized into four distinct regions based on the bond percentage, each represented by a different color: 0–5\% (GC or graphite structure), 5–80\% (a-C structure), 80–88\% (ta-C structure), and 88–100\% (a-D structure).}
	\label{2900K}
\end{figure}

\vskip 3cm

\section{Ensuring the Reliability of Critical Pressure Determination}
\begin{figure}[H]
	\centering
	\includegraphics[width=0.85\textwidth]{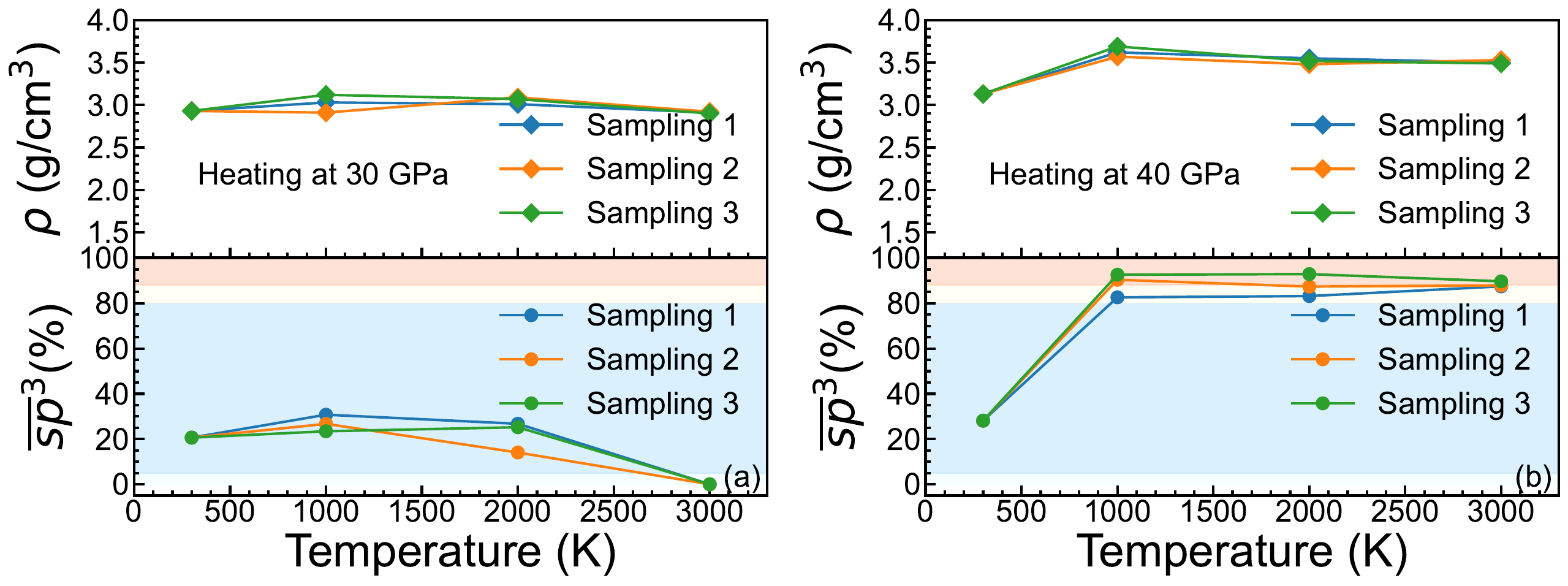}
	\caption{The average density (upper panel) and the percentage of sp\textsuperscript{3} bonds (lower panel) as a function of temperature for another three heating simulations conducted at 30~GPa (a) and 40~GPa (b). The percentage of sp\textsuperscript{3} bonds is categorized into four distinct regions based on the bond percentage, each represented by a different color: 0–5\% (GC or graphite structure), 5–80\% (a-C structure), 80–88\% (ta-C structure), and 88–100\% (a-D structure).}
	\label{sampling}
\end{figure}